\documentclass[aps,prb,floatfix,superscriptaddress,noshowpacs,twocolumn]{revtex4-2}
\usepackage{subcaption}
\usepackage{graphicx,amsmath,amssymb,epstopdf,wasysym,ulem}
\usepackage[space]{grffile}
\usepackage{hyperref}
\usepackage{bm}
\usepackage[dvipsnames]{xcolor}
\usepackage{braket,amsfonts}
\usepackage{dsfont}
\usepackage{float} 
\allowdisplaybreaks

\newcommand{\be}{\begin{equation}}
\newcommand{\ee}{\end{equation}}
\newcommand{\bga}{\begin{gather}}
\newcommand{\ega}{\end{gather}}
\newcommand{\bea}{\begin{eqnarray}}
\newcommand{\eea}{\end{eqnarray}}
\newcommand{\dagga}{{\phantom{\dagger}}}
\newcommand{\bR}{\mathbf{R}}

\newcommand{\bq}{\mathbf{q}}

\newcommand{\bp}{\mathbf{p}}

\newcommand{\bx}{\mathbf{x}}

\newcommand{\br}{\mathbf{r}}
\newcommand{\bu}{\mathbf{u}}

\newcommand{\dis}{\displaystyle}

\newcommand{\fract}[2]{\frac{\dis \;#1\;}{\dis \;#2\;}}

\newcommand{\eqn}[1]{(\ref{#1})}

\newcommand{\ep}{{\epsilon}}

\newcommand{\bw}{\begin{widetext}}
\newcommand{\ew}{\end{widetext}}
\newenvironment{eqs}%
{\begin{equation} \begin{aligned}}%
{\end{aligned} \end{equation} }
\newcommand{\beal}{\begin{eqs}}
\newcommand{\eal}{\end{eqs}}
\newcommand{\bd}[1]{{\boldsymbol{#1}}}
\newcommand{\esp}[1]{\text{e}^{#1}}
\newcommand{\bealn}{\beal\nonumber}
\newcommand{\coloneq}{\mathrel{\mathop:}\mathrel{\mkern-1.2mu}=}

\begin{document}

\title{How chiral vibrations drive molecular rotation}

\author{Ivan Pasqua} 
\affiliation{International School for
  Advanced Studies (SISSA), Via Bonomea
  265, I-34136 Trieste, Italy} 
\author{Gregorio Staffieri} 
\email{Contact author: gstaffie@sissa.it}
\affiliation{International School for
  Advanced Studies (SISSA), Via Bonomea
  265, I-34136 Trieste, Italy}
\thanks{PROVA}  
\author{Michele Fabrizio} 
\affiliation{International School for
  Advanced Studies (SISSA), Via Bonomea
  265, I-34136 Trieste, Italy} 



\begin{abstract}
We analyze two simple model planar molecules: an ionic molecule with $D_3$ symmetry and a covalent molecule with $D_6$ symmetry. Both symmetries allow the existence of chiral molecular orbitals and normal modes that are coupled to each other in a Jahn-Teller manner, invariant under $U(1)$ symmetry with generator a pseudo angular momentum. 
In the ionic molecule, the chiral mode possesses an electric dipole but lacks physical angular momentum, whereas, in the covalent molecule, the situation is reversed. In spite of that,
we show that in both cases the chiral modes can be excited by a circularly polarized light and are subsequently able to induce rotational motion of the entire molecule.

\end{abstract}

\maketitle
\section*{Introduction}

The Einstein-de Haas (EdH) effect \cite{Einstein-deHaas} and its reciprocal, the Barnett effect \cite{Bernett-PR1915}, 
were discovered over a century ago, but are currently experiencing a revival of interest \cite{Niu-PRL2014,Garanin-PRB2015,Dornes-Nature2019,Rembert-PRB2020,Garanin-PRB2021,Tauchert-Nature2022}, especially after the prediction \cite{Niu-PRL2015,Hamada-PRL2018,Zhong-PRB2023}, 
and possible observation \cite{zhang-ArXiv2024} that chiral phonons may play the same role of the electron spins in the original experiments \cite{Einstein-deHaas,Bernett-PR1915}, 
see \cite{Zhang-NanoLett2024} for a recent review. \\
While the ultimate explanation of the EdH effect is the conservation of total angular momentum,
the precise mechanisms that enable the transfer of the angular momentum carried by microscopic internal degrees of freedom to the macroscopic rigid body rotation are considerably more challenging to elucidate. This stems primarily from the fact that commonly employed model Hamiltonians do not explicitly incorporate global degrees of freedom, such as the sample angle of rotation.  
Furthermore, the EdH effect caused by the excitations of specific phonon modes encounters an additional challenge. In most cases, these phonons only carry a pseudo angular momentum \cite{Streib-PRB2021}, and it remains unclear how the latter can be converted into a physical angular momentum.  
It has also been noted \cite{Gregory-PRB2024} that a faithful description of experiments that detect evidence of phonon magnetic moments necessitates degenerate chiral phonons coupled to degenerate electronic orbitals, thereby realizing a form of Jahn-Teller effect that is in fact associated with the conservation of a pseudo angular momentum distinct from the physical one. \\

\noindent
Based on these observations, in this work we aim to clarify these questions by introducing and analyzing two toy models that describe two distinct types of planar molecules: an ionic and a covalent one, exemplified by metal trifluoride MF$_3$ (M=Al, Sc, Y),  and by benzene, C$_6$H$_6$, respectively. These models possess chiral degenerate normal modes and degenerate molecular orbitals, thereby accurately capturing the physical phenomena discussed above. 
The choice of ionic and covalent molecules is motivated by their realization of two distinct types of Jahn-Teller coupling.
 \\
In the ionic molecule discussed in Sec.~\ref{$D_3$ symmetric ionic molecule}, 
the Jahn-Teller effect originates from the dependence on the ion displacement of the crystal field experienced by the electrons, and explicitly involves the global angle of rotation of the molecule, as shown in Sec.~\ref{Jahn-Teller coupling triangle}.     
Although this aspect is often overlooked, it is crucial for understanding how the pseudo-angular momentum associated with the Jahn-Teller effect interacts with the physical angular momentum. Through this interaction, as demonstrated in two hypothetical experiments, a circularly polarized light pulse induces the rotation of the entire molecule, as presented in Sec.~\ref{Hypothetical experiment triangle}.
 \\
Conversely, in the covalent case discussed in Sec.~\ref{$D_6$ symmetric covalent molecule}, the Jahn-Teller effect arises from the atomic displacements 
affecting the electron covalent bonding, Sec.~\ref{Electron Hamiltonian},
and does not directly involve the molecule's rotation angle, as shown in Sec.~\ref{Coupling to the normal modes}. However, the chiral normal modes now possess a physical angular momentum, even though, lacking an electric dipole, they cannot directly couple to an electromagnetic field. Nonetheless, through the virtual photoexcitation of an optically active particle-hole pair, that subsequently de-excite via Jahn-Teller coupling, the electromagnetic field can couple to the chiral modes, 
making them detectable in the infrared absorption spectrum, as discussed in 
sections \ref{Coupling to a circularly polarized electromagnetic field} and 
\ref{Hypothetical experiment benzene}.  A similar phenomenon occurs, for instance, in doped C$_{60}$ and explains the optical absorption at the frequency of the T$_{1u}$ modes \cite{Rice&Choi}. In this context, we also propose in Sec.~\ref{Hypothetical experiment benzene} a hypothetical 
experiment where a circularly polarized light in resonance with the chiral normal modes can induce rotational motion in the molecule.

\begin{figure}[t]
\vspace{0.2cm}
\centerline{\includegraphics[width=0.3\textwidth]{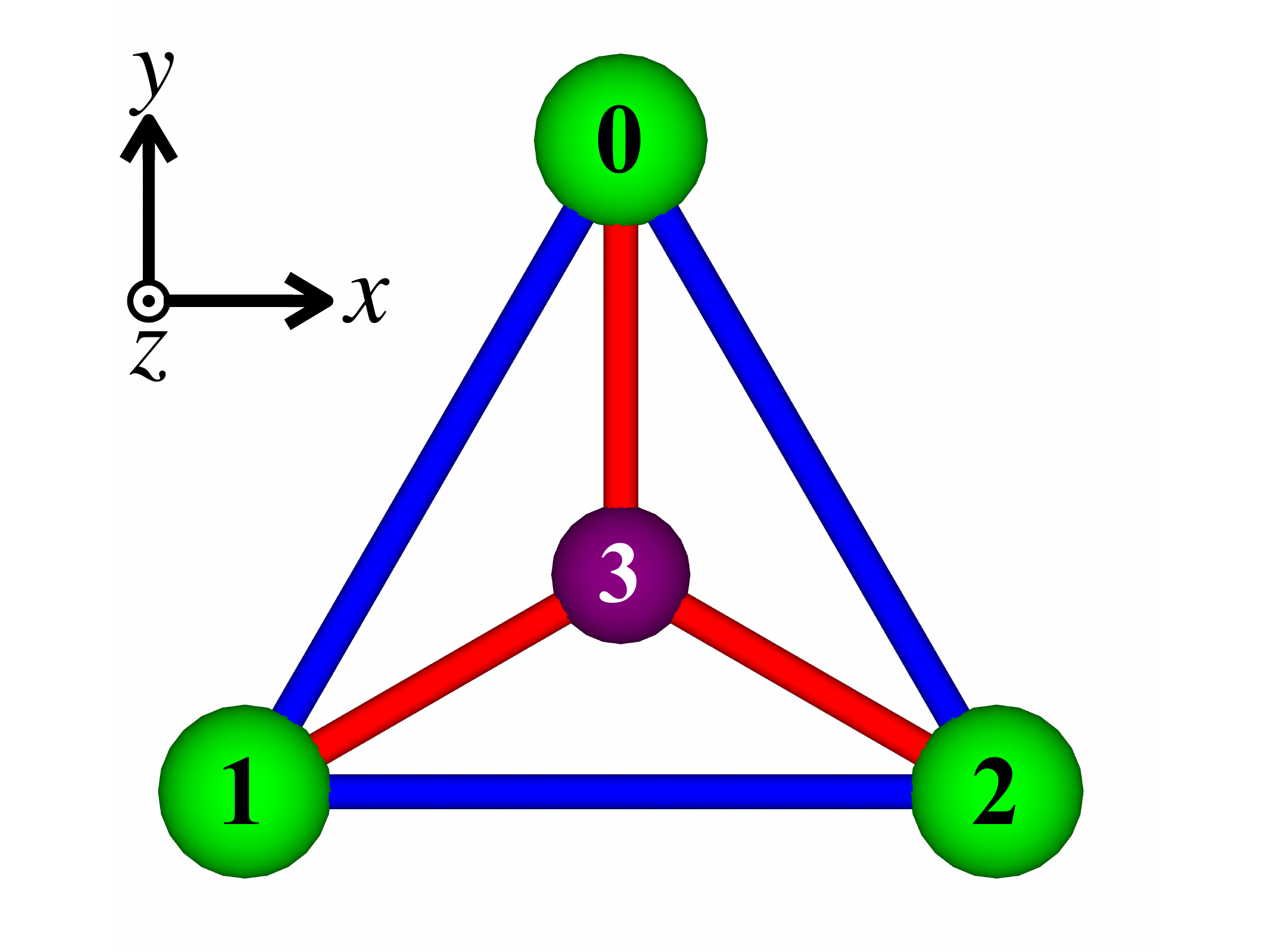}}
\caption{Toy planar molecule with $D_3$ symmetry. The bonds represent springs with spring constants 
$K$, the blue bonds, and $\gamma K$, the red ones.  } 
\label{NF3}
\end{figure}

\section{$D_3$ symmetric ionic molecule}
\label{$D_3$ symmetric ionic molecule}
The first toy model we study refers to an ionic molecule that exhibits $D_3$ planar symmetry, depicted schematically in Fig.~\ref{NF3}. 
Even though we do not intend to faithfully describe any real molecule, we could nonetheless imagine that the physics we are going to describe might be pertinent, e.g., to metal trifluoride MF$_3$ (M=Al, Sc, Y), which is indeed a planar ionic molecule. \\
We adopt a reference frame in which the position of the central ion, numbered as 3 and with purple color in Fig.~\ref{NF3}, is fixed, but the molecule can freely rotate around that position, with rotation angle $\phi$. The three external ions, numbered from 0 to 2 and drawn in green, have equal mass $M$ and are symmetrically located around the central one with an equilibrium distance that we take as our unit of length. From now on, we disregard the out-of-plane ionic motion. Consequently, the point group is the two-dimensional $D_3$ that includes, besides the identity $E$, the threefold rotation $C_3$ around atom 3 and the twofold rotation $C'_2$ around the $y$-axis and its equivalent axes by $C_3$ symmetry.
The character 
and product tables of the irreducible representations (irreps) are shown in Table~\ref{Table D3}.
\begin{table}
\be\nonumber
\begin{array}{|c|c|c|c|}\hline
\bd{D_3} & E & C_3 & C'_2 \\ \hline
A_1 & 1 & 1 & 1 \\ \hline
A_2 & 1 & 1 & -1 \\ \hline
E & 2 & -1 & 0 \\ \hline
\end{array}
\quad 
\begin{array}{|c||c|c|c|}\hline
~ & A_1 & A_2 & E \\ \hline\hline
A_1 & A_1 & A_2 & E \\ \hline
A_2 & A_2 & A_1 & E \\ \hline
E & E & E & A_1\oplus\left[A_2\right]\oplus E \\ \hline
\end{array}
\ee
\caption{Character table, left, and product table, right, of the wallpaper group $D_3$. The 
antisymmetric product of two identical irreducible representations is indicated by square brackets. }
\label{Table D3}
\end{table} 
Following the Appendix, we parametrize the positions of the 
green atoms, $n=0,1,2$, as 
\beal
\br_n &= \bR_n + \bx_n = \big(1+\alpha_n)\,\bR_n + \beta_n\,\bd{z}\wedge\bR_n\,,
\label{r-n triangle}
\eal  
with $\bd{z}$ the unit vector perpendicular to the plane of the molecule, 
\beal
\bR_n &= C_{n2\pi/3}\big(\bR_0\big)\,,& \bR_0&=(0,1)\,,
\label{R-n triangle}
\eal
the equilibrium positions, with $C_\theta$ representing the rotation by an angle $\theta$ around the central atom, and $\bx_n$ the $n$-th ion displacement. 
We further write, see \eqn{Fourier appx}, 
\beal
\begin{pmatrix}
\alpha_n\\
\beta_n
\end{pmatrix} 
&= \fract{1}{\sqrt{3}}\sum_{\ell=-1}^1\,
\esp{i k_\ell n}\;\bq_\ell\,,
\label{Fourier triangle}
\eal
where $k_\ell = 2\pi\ell/3$, $\ell=-1,0,+1$. As discussed in the Appendix, to avoid double counting the variable $\phi$ that describes the molecular rotation with respect to a fixed reference frame, the vector $\bq_\ell$ for $\ell=0$ has only one component that differs from zero, i.e., 
\bealn
\bq_0 &= \begin{pmatrix}
q_{A_1}\\
0
\end{pmatrix} \,,
\eal
and describes the $A_1$ breathing mode. On the contrary, $\bq_\ell^\dagga=\bq_{-\ell}^*$ for $\ell=\pm1$ have both components finite and transform as the two-dimensional irrep $E$.
The atomic Hamiltonian has the general form, see \eqn{H DN}, 
\beal
H_\text{at} &= \fract{\;\big(p_\phi- \text{L}_\text{vib}\big)^2\;}{2I}+ 
\fract{\;p_{A_1}^2\;}{2M} \\
&\qquad + \fract{1}{2M}\,\sum_{\ell\not=0}\,\bp_{\ell}\cdot\bp_{-\ell} + V\big(\{\br_n\}\big)\,,
\label{H triangle 0}
\eal 
where $p_\phi$ is the momentum conjugate to $\phi$ and equals the total angular momentum 
$\text{L}$, while $p_{A_1}$ is conjugate to $q_{A_1}$ and, for $\ell=\pm 1$, $\bp_{-\ell}$ conjugate to $\bq_\ell$. The operator
\be
\text{L}_\text{vib} = \sum_{\ell=\pm 1}\,\bq_\ell\wedge\bp_{-\ell}\cdot\bd{z}\,,
\label{L vib triangle 0}
\ee
is the contribution of the vibrational modes to the angular momentum, sometimes called the 
\textit{spin} of the phonons, 
and $I = M\,\left(\sqrt{3\,} + q_{A_1}\right)^2$ the moment of inertia. \\
The eigenstates of \eqn{H triangle 0} can be classified in terms of the 
irreps $A_1$ and $E$ in Table~\ref{Table D3}. In particular, the operator 
\eqn{L vib triangle 0} transforms as $A_2$, which implies that it couples eigenstates with symmetry $E$, see the product rules in Table~\ref{Table D3}. 

\subsection{Inter atomic potential in the harmonic approximation}
\begin{figure}[ht]
\centerline{\includegraphics[width=0.3\textwidth]{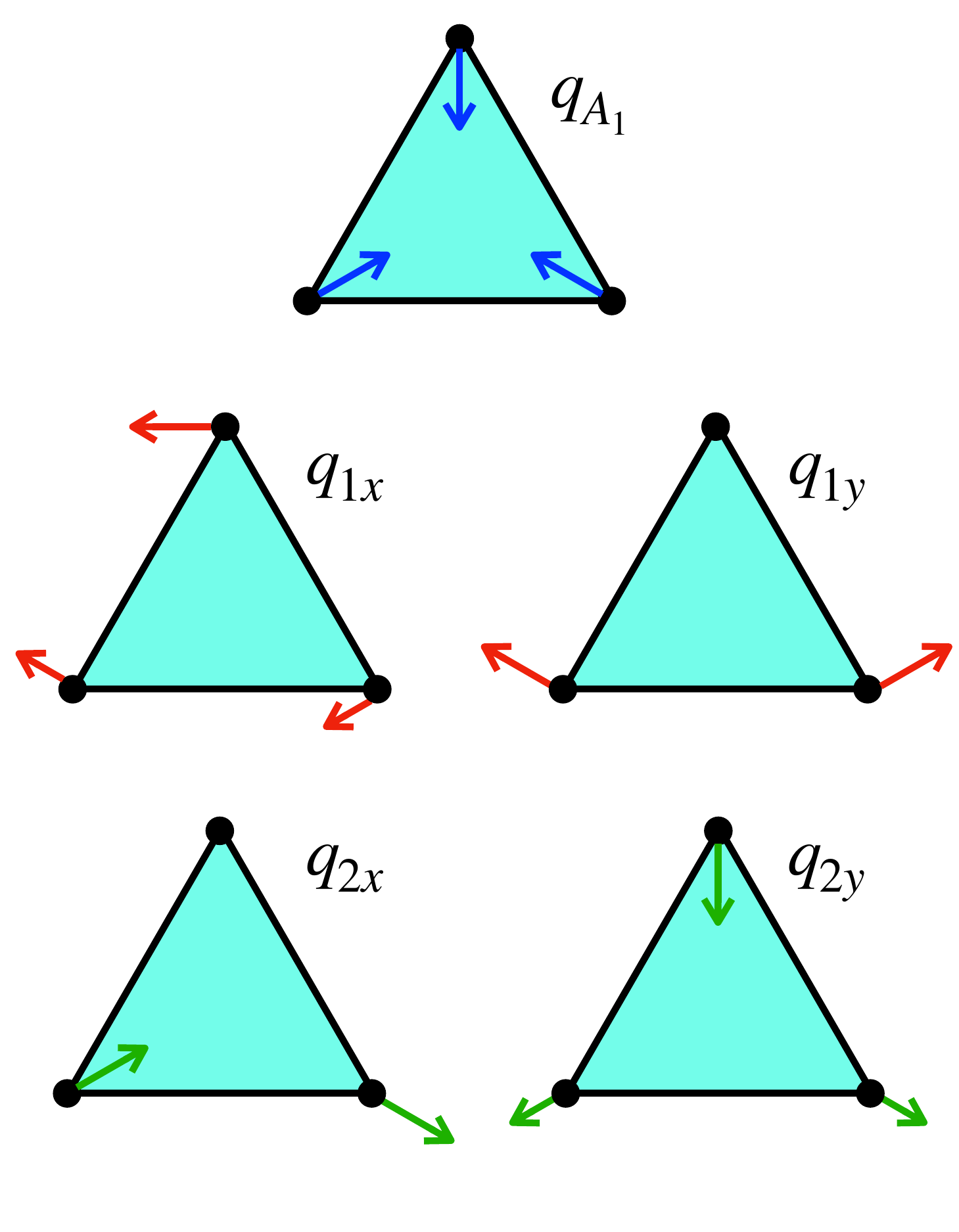}}
\vspace{-0.2cm}
\caption{Sketch of the normal modes of the molecule in Fig.~\ref{NF3} in the limit $\gamma\gg 1$.  } 
\label{NF3-modes}
\end{figure}

To simplify the analysis, we adopt the harmonic approximation for the interatomic potential, which is quite valid in AlF$_3$ \cite{Pak-JCP1997,Nejad-PCCP2020} as well as in ScF$_3$ and Y$_3$ \cite{Solomik-JSC2000}.
Specifically, we assume that the blue and red bonds in Fig.~\ref{NF3} have, respectively, spring constants $K$ and $\gamma\,K$, with $\gamma\gg 1$ \cite{Pak-JCP1997}, consistently with the ionic nature of the red bonds. Therefore, using the parametrization \eqn{r-n triangle}, $V\big(\{\br_n\}\big)$ in the harmonic approximation is simply 
\beal
&V\big(\{\br_n\}\big) = \fract{\gamma K}{2}\,\sum_{n=0}^2\, \alpha_n^2 \\
&\quad + \fract{K}{8}\,\sum_{n=0}^2\,
\bigg(\sqrt{3}\,\big(\alpha_{n+1}+\alpha_n\big)
+\big(\beta_{n+1}-\beta_n\big)\bigg)^2\,.
\label{potential triangle}
\eal
The normal modes are obtained by diagonalizing the dynamical matrix. This is readily accomplished using \eqn{Fourier triangle}, leading to the eigenvalue equations for the dynamical matrix 
\beal
\lambda_\ell^2\,\bu_\ell &= 
\begin{pmatrix}
\varepsilon_\ell + A_\ell & i B_\ell\\
-i B_\ell & \varepsilon_\ell - A_\ell
\end{pmatrix}\,\bu_\ell\,,
\label{eigenvalue eq triangle 2}
\eal
where $\lambda^2_\ell$ are the eigenvalues in units of $K$, $B_\ell = -B_{-\ell}=\sin k_\ell\,\sqrt{3}/2$ and 
\bealn
\varepsilon_\ell &= \fract{\gamma+2+\cos k_\ell}{2}\;,&
A_\ell &= \fract{\gamma+1+2\cos k_\ell}{2}\;.
\eal
The eigenvalues are therefore
\be
\lambda^2_{1\ell} = \ep_\ell -\sqrt{A_\ell^2+B_\ell^2\;}\;,\quad
\lambda^2_{2\ell} = \ep_\ell +\sqrt{A_\ell^2+B_\ell^2\;}\;,
\ee
and the corresponding eigenvectors read
\bealn
\bu_{1\ell} &= \begin{pmatrix}
-i\sin\theta_\ell\\
\cos\theta_\ell
\end{pmatrix}\,,&
\bu_{2\ell} &= \begin{pmatrix}
\cos\theta_\ell\\
-i\sin\theta_\ell
\end{pmatrix}\,,
\eal
where
\bealn
\theta_\ell &=-\theta_{-\ell} = \fract{1}{2}\,\tan^{-1}\fract{B_\ell}{A_\ell}\;.
\eal
For $\ell=0$, where $\theta_0=0$, only the mode 2 with eigenvalue $\lambda^2_{20}=\lambda^2_{A_1}=3+\gamma$ must be considered, and corresponds to the $A_1$ breathing mode 
with normal mode coordinate $q_{A_1}$. The modes with $\ell=+1$ are characterized by  
\beal
&\theta_1 = \theta=\fract{1}{2}\,\tan^{-1}\fract{3}{2\gamma}
\xrightarrow[\gamma\gg 1]{} \fract{3}{4\gamma}\ll 1
\;,\\
&\lambda^2_{1+1}=\lambda_1^2 \xrightarrow[\gamma\gg 1]{} \fract{5}{4}\;,\qquad 
\lambda^2_{2+1}=\lambda_2^2\xrightarrow[\gamma\gg 1]{}\gamma+\fract{1}{4}\;,
\label{theta triangle}
\eal
and are degenerate with the $\ell=-1$ ones that have $\theta_{-1}=-\theta$. For $\ell=\pm 1$ we introduce 
the normal mode coordinates, $q_{1\ell}$ and $q_{2\ell}$, and momenta, $p_{1\ell}$ and $p_{2\ell}$. Correspondingly, the coordinate, $\bq_\ell$, and momentum, $\bp_{\ell}$, operators in 
\eqn{H triangle 0} and \eqn{L vib triangle 0} become
\be\nonumber
\bq_\ell = q_{1\ell}\,\bu_{1\ell} + q_{2\ell}\,\bu_{2\ell}\,,\quad
\bp_{\ell} = p_{1\ell}\,\bu_{1\ell} + p_{2\ell}\,\bu_{2\ell}\,.
\ee
We find more convenient to introduce real normal mode coordinates  
through the canonical transformation 
\bealn
q_{a\pm 1} &=\fract{1}{\sqrt{2}}\,\big(q_{ax}\pm i\, q_{ay}\big)\,,& a&=1,2\,,
\eal
and similarly for the conjugate momenta. In terms of the new conjugate operators, Eq.~\eqn{L vib triangle 0} reads
\beal
\text{L}_\text{vib} &= -\sin2\theta\,\Big(\bq_1\wedge\bp_1 -\bq_2\wedge\bp_2\Big)\\
&\qquad -\cos2\theta\,\Big(\bq_1\wedge\bp_2 +\bq_2\wedge\bp_1\Big)\,,
\label{L-vib triangle}
\eal
where $\bq_a$, $a=1,2$, is the vector with components $q_{ax}$ and $q_{ay}$. 
The Hamiltonian \eqn{H triangle 0} in the harmonic approximation 
has the simple expression 
\beal
H_\text{at} &= \fract{ 1}{\;2I\;}\,\Big(p_\phi - \text{L}_\text{vib}\Big)^2 + 
\fract{1}{2M}\,\sum_j\,p_j^2 \\
&\qquad + \fract{K}{2}\,\sum_j\,\lambda_j^2\,q_j^2\,,
\label{Ham triangle}
\eal
where $j=A_1,1x,1y,2x,2y$ and, we recall, 
\be
I = M\,\big(\sqrt{3\,}+q_{A_1}\big)^2\,.
\label{I triangle}
\ee
The normal modes are shown in Fig.~\ref{NF3-modes} for large $\gamma$. 

\subsection{Jahn-Teller coupling}
\label{Jahn-Teller coupling triangle}
The electron molecular orbitals (MOs) can also be classified based on the irreducible representations in Table~\ref{Table D3}, or, more precisely, the irreps of the point group $D_{3h}$. Let us here consider a MO that transforms like the two-dimensional irrep $E$, $E'\sim (x,y)$ in $D_{3h}$, and explore the Jahn-Teller coupling with the ionic displacements. For simplicity, we restrict our analysis to the linear order in the ionic displacement. Later, we will discuss the effects of the second-order term. The electrons on the $E$-type MO feel a crystal field potential 
\beal
U &= \sum_{n=0}^2\,U\big(|\br-\br_n|\big) = \sum_{n=0}^2\,U\big(\left|\br-\bR_n\right|\big)\\
&\quad +\sum_{n=0}^2\,\Big\{U\big(\left|\br-\bR_n-\bx_n\right|\big)-
U\big(\left|\br-\bR_n\right|\big)\Big\}\\
&= U_0 + U_\text{JT}\,,
\eal
where $\br$ is the electron coordinate, $U_0$ the potential due to the atoms in their equilibrium positions $\bR_n$, while $U_\text{JT}$ is the correction caused by the atomic displacements, which starts linear in $\bx_n$. It follows that $U_0$ is invariant under $D_3$, and therefore cannot split the MO. 
On the contrary, $U_\text{JT}$ can lift the MO twofold degeneracy by the Jahn-Teller effect. 
To find an explicit expression of $U_\text{JT}$, we assume the MO localized on the central atom, 3 in Fig.~\ref{NF3}, 
so that $|r| \ll |\bR_n|$, and expand $U_\text{JT}$ up to second order in $\br$ and at first order 
in the atomic displacements. The first order correction in $\br$ gives no contribution since the 
two orbitals have the same odd parity. The calculation of the second order correction 
in $\br=r\,\left(\cos\phi_e,\sin\phi_e\right)$ is simple but tedious, therefore we present here just the final result
\beal
U_\text{JT} &= \fract{3r^2}{4}\bigg\{g_1\Big(\cos2\big(\phi-\phi_e\big)\,q_{1y}\\
&\qquad \qquad\qquad\qquad + \sin 2\big(\phi-\phi_e\big)\,q_{1x}\Big)\\
&\qquad\qquad\quad  + g_2\,\Big(\cos2\big(\phi-\phi_e\big)\,q_{2y}\\
&\qquad \qquad\qquad\qquad\quad
+ \sin 2\big(\phi-\phi_e\big)\,q_{2x}\Big) \bigg\}\,,
\label{HJT triangle}
\eal
where
\bealn
g_1 &= U_1\,\sin\theta + U_2\,\cos\theta\,,& g_2 &= U_1\,\cos\theta - U_2\,\sin\theta\,,
\eal
and  
\bealn
U_1 &= -\fract{1}{2}\big(\partial U- \partial^2 U+ \partial^3 U\big)\,,&
U_2 &= -\big(\partial U- \partial^2 U\big)\,,
\eal
with $\partial^n U$ the $n$-th derivative of $U\big(|\br-\bR_n|\big)$ calculated 
at $\br=\bd{0}$, thus at $|\br-\bR_n|=|\bR_n|=1$. The potential \eqn{HJT triangle} projected 
onto the MO subspace, with orbitals $p_x(\br)\sim \psi(r)\,\cos\phi_e$ and 
$p_y(\br)\sim \psi(r)\,\sin\phi_e$, realizes a standard $e\times E$ Jahn-Teller effect 
with a peculiar property. Indeed, the full Hamiltonian, including $U_\text{JT}$ in \eqn{HJT triangle}, preserves the operator
\beal
\text{L}_\text{TOT} &= \text{L}+\text{L}_e = -i\,\bigg(
\fract{\partial}{\partial \phi} + \fract{\partial}{\partial \phi_e}\bigg)\,,
\eal
which we can legitimately regard as the total, electrons plus ions, angular momentum. A posteriori, 
this result is in no way surprising, since the total Hamiltonian must be invariant under global planar rotations. However, this property is not often emphasized and, as we are going to show, leads to 
interesting phenomena.

\subsection{Hypothetical experiments}
\label{Hypothetical experiment triangle}
Here and in Section~\ref{Hypothetical experiment benzene}, we explore how a light pulse circularly polarized in the plane of the molecule can induce molecular rotations. We will consider this possibility in a highly idealized setup: an isolated molecule struck by the laser pulse. A more realistic scenario could involve molecules in the gas phase within a cylindrical vessel that can rotate around its axis, and a light pulse with circular polarization perpendicular to that axis. In this case, we must also account for additional complications, such as intermolecular interactions, the interaction between the molecules and the vessel wall, and the thermally randomized orientation of the molecule planes relative to the light polarization.  
Although such an experiment can be envisioned, we cannot predict whether the desired resulting rotation of the vessel would be measurable. Therefore, we restrict our analysis to the somewhat unrealistic setup of an isolated molecule.  \\
In the specific case of the molecule in Fig.~\ref{NF3}, since $\gamma\gg 1$, the lowest energy normal mode is the mode $\bq_1$ with symmetry $E$ and frequency 
$\omega_{1} \equiv \omega_0 \simeq \sqrt{5K/4M\,}$. It describes displacements perpendicular to the equilibrium positions $\bR_n$, the $q_{1x}$ and $q_{1y}$ modes of Fig.~\ref{NF3-modes}. In what follows, we neglect all other higher energy modes, so that the moment of inertia \eqn{I triangle} becomes constant, $I\simeq 3M$.   
Since $\gamma\gg 1$ also implies $\theta\simeq 0$, see \eqn{theta triangle}, the contribution of the normal mode 1 to the angular momentum in \eqn{L-vib triangle} can be neglected. We observe that the off-diagonal term in \eqn{L-vib triangle}, which remains finite for $\theta\to 0$, can still provide mode 1 with an angular momentum at second order perturbation theory. However, such term for $\gamma\gg 1$ is of the same order as the diagonal one $\propto \sin2\theta$, and thus equally negligible. We mention that there are instead circumstances in which off-diagonal terms in $\text{L}_\text{vib}$ may play the major role 
\cite{Zhong-PRB2023}.\\
Focusing, as before, on an $E$-type MO, the total Hamiltonian, sum of \eqn{Ham triangle} and \eqn{HJT triangle}, can therefore be written as 
\beal
H &\simeq H_\text{at} + U_\text{JT} = \fract{\;p_\phi^2\;}{2I}
+ \fract{\omega_0}{2}\,\Big(\bp^2 + \bq^2\Big)\\
&\; -g\,\sum_\sigma\,\bigg\{ 
q_{x}\,\Psi_\sigma^\dagger\,\Big(\cos2\phi\,\tau_3 + \sin2\phi\,\tau_1\Big)\,\Psi^\dagga_\sigma\\
&\qquad\qquad\;
+ q_{y}\,\Psi^\dagger_\sigma\,\Big(\sin2\phi\,\tau_3 - \cos2\phi\,\tau_1\Big)\,\Psi^\dagga_\sigma
\bigg\}\,,
\label{H total triangle}
\eal 
where $\bq$ and $\bp$ are dimensionless conjugate variables with components $q_a$ and $p_a$, $a=x,y$, defined as 
\bealn
q_a &= \sqrt{M\omega_0\;}\,q_{1a}\,,& p_a &=\sqrt{\fract{1}{M\omega_0}\;}\,p_{1a}\,.
\eal
The spinor $\Psi^\dagga_\sigma$ in \eqn{H total triangle} is defined as   
\bealn
\Psi^\dagga_\sigma &= \begin{pmatrix}
c^\dagga_{x\sigma}\\
c^\dagga_{y\sigma}
\end{pmatrix}\,,
\eal
with components the annihilation operators of the MO orbitals $p_x$ and $p_y$ with spin 
$\sigma$, and $\tau_\alpha$, $\alpha=1,2,3$, the Pauli matrices that act in the orbital space. 
In this representation, the electron angular momentum reads
\bealn
\text{L}_e &= \sum_\sigma\,\Psi_\sigma^\dagger\,\tau_2\,\Psi_\sigma^\dagga\,,
\eal
and $\text{L}_\text{TOT}= p_\phi+\text{L}_e$ is conserved. We apply a $\pi/2$ rotation around $\tau_1$, so that $\tau_2\to\tau_3$, $\tau_3\to-\tau_2$ and thus  
\bealn
\Psi^\dagga_\sigma &\to 
\begin{pmatrix} c^\dagga_{+1\sigma}\\ c^\dagga_{-1\sigma}\end{pmatrix}\,,& 
\text{L}_e &\to \sum_\sigma\,\Psi_\sigma^\dagger\,\tau_3\,\Psi_\sigma^\dagga\,.
\eal
Moreover, we define
\bealn
q_x &= -\fract{i}{2}\big(a^\dagga_+ - a^\dagger_+ - a^\dagga_- + a^\dagger_-\big)\,,\\
q_y &= -\fract{1}{2}\big(a^\dagga_+ + a^\dagger_+ + a^\dagga_- + a^\dagger_-\big)\,,
\eal
and accordingly $p_x$ and $p_y$,  
where $a^\dagga_{\pm}$ and $a^\dagger_{\pm}$ are bosonic annihilation and creation operators, respectively, so that 
\beal
H &= \fract{\;p_\phi^2\;}{2I} + \omega_0\,\Big(n_++n_-+1\Big)\\
&\quad -g\,\sum_\sigma\,\bigg\{ \big(a^\dagga_{+} + a^\dagger_{-}\big)\, 
\Psi_\sigma^\dagger\;\esp{-i2\phi}\;\tau_+\;\Psi^\dagga_\sigma\\
&\qquad\qquad\qquad
+\big(a^\dagger_{+} + a^\dagga_{-}\big)\, \Psi^\dagger_\sigma\;\esp{i2\phi}\;\tau_-\;
\Psi^\dagger_\sigma
\bigg\}\,,
\label{H total triangle bis}
\eal 
with $n_\pm = a^\dagger_\pm\,a^\dagga_\pm$, 
which further emphasizes the existence of another conserved quantity of $H$, specifically, 
\beal
\text{J} &= n_+ - n_- + \fract{1}{2}\,\text{L}_e\,,
\label{pseudo angular momentum triangle}
\eal
which plays the role of a pseudo angular momentum. \\
We note that $\text{J}$ assumes integer values when the number of electrons on the MO is even, and half-integer values when the number of electrons is odd, which is the case we are interested in.  
The latter property is unique to a linear Jahn-Teller coupling. In the presence of a second-order term in the atomic displacement, and as its coupling strength increases, the pseudo angular momentum $\text{J}$ at odd occupation of the MO quite abruptly crossovers from being quantized in half-integer values to being quantized in integer values 
\cite{Zwanziger&Grant-1987}, the 1/2 in \eqn{pseudo angular momentum triangle} simply replaced by 1. Nevertheless, a pseudo angular momentum can still be defined and differs from the physical one for the simple fact that the mode 1 carries no angular momentum. 
Therefore, despite our focus on the linear order term, which notably simplifies the calculations, we anticipate no qualitative changes in the presence of a second-order correction.   
\\

\begin{figure}[t]
\vspace{0.2cm}
\centerline{\includegraphics[width=0.5\textwidth]{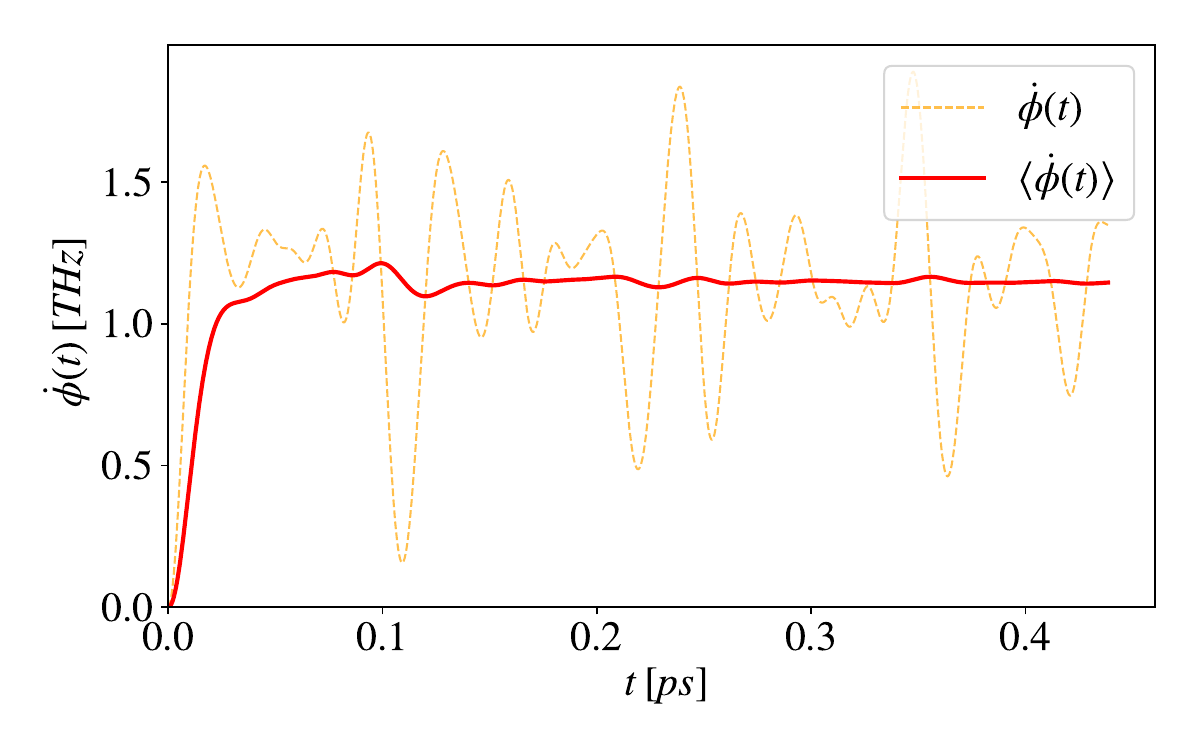}}
\caption{Time evolution of the angular velocity $\dot{\phi}(t) =  p_\phi(t) / I$, and its time-average $\langle \dot{\phi}(t)\rangle$, obtained by integrating the Lanczos chain defined in \eqn{Lanczos chain}. The undamped oscillations over time result from the exchange of angular momentum between the molecule and the electron in the singly occupied $E$-type molecular orbital during the unitary evolution. The chosen parameters correspond to those of AlF$_{3}$, with 
$\omega_0 = \nu_4 \simeq 240~\text{cm}^{-1}$ \cite{Pak-JCP1997}, 
a moment of inertia $I \simeq 36 $ in units of $\hbar ^{2} / \omega_{0}$. The Jahn-Teller coupling $g$ obtained through \eqn{HJT triangle} is $1.8\,\omega_0$. } 
\label{FigureToy1}
\end{figure}

\noindent
Although our primary objective is to demonstrate how chiral normal modes can be excited by light and subsequently induce molecular rotation, we begin by presenting a different hypothetical experimental scenario. In this scenario, light is employed to generate an electronic excitation, and we show how, through the Jahn-Teller coupling, this excitation is transformed into a molecule's rotation.  \\
We assume that the above $E$-type MO is empty in the ground state, which has vanishing total angular momentum. However, we can envision a scenario where, at $t=0$, the molecule is in an excited state characterized by one electron in that molecular orbital with $\text{L}_e=+1$, yet no excited bosons. This possibility could arise, for instance, if an ultrafast circularly-polarized light pulse were to transfer one electron from an occupied MO 
of symmetry $A_1$ to the unoccupied MO of symmetry $E$.
It follows that the initial values of total and pseudo angular momenta are $\text{L}_\text{TOT}=1$ and $\text{J}=1/2$, and they remain so under the unitary evolution with Hamiltonian 
\eqn{H total triangle bis}. Along this evolution, $\text{L}_e$ jumps between +1 and -1  
and, correspondingly, $n_+-n_-$ between 0 and +1, so as to maintain $\text{J}=1/2$ constant. 
Therefore, the contribution of the rigid body rotation to the angular momentum
\bealn
p_\phi= I\,\dot{\phi}=\text{L}_\text{TOT} -  \text{L}_e = 1 - \text{L}_e\,,
\eal
jumps between 0 and 2, namely the molecule acquires an angular acceleration after 
the purely electronic excitation.\\ 
To simulate the actual dynamics,  
we construct the Lanczos chain where the first site $\ell=1$ is the initial state, i.e., 
$\ket{1} = c^\dagger_{+1}\ket{0}$, with $\ket{0}$ the electron and boson vacuum. In this way 
we can formally rewrite \eqn{H total triangle bis} as
\beal
H &= \sum_{\ell\geq 1}\,t_\ell\,\Big(\ket{\ell+1}\bra{\ell}
+\ket{\ell}\bra{\ell+1}\Big)\\
&\qquad +\sum_{\ell\geq 1}\,\ep_\ell\,\ket{\ell}\bra{\ell}\,,
\label{Lanczos chain}
\eal
where the parameters can be easily derived and read  
\bealn
t_\ell &= 
\begin{cases}
-g\,\sqrt{\ell/2}\;,& \ell=\text{even}\,,\\
-g\,\sqrt{(\ell+1)/2}\;,& \ell=\text{odd}\,,
\end{cases}\\
\ep_\ell &= \omega_0\, \ell + \begin{cases}
2/I\;,& \ell=\text{even}\,,\\
0\;,& \ell=\text{odd}\,,
\end{cases}
\eal
while the sites of the chain correspond to the wavefunctions
\bealn
&\ket{\ell=2n+1} = \fract{1}{\sqrt{2\pi}}\;
\fract{a^\dagger_+{^{n}}\,a^\dagger_-{^{n}}}{n!}\,\, c^\dagger_{+1}\ket{0}\,,\\
&\ket{\ell=2n+2} = \fract{\esp{2i\phi}}{\sqrt{2\pi}}\;
\fract{a^\dagger_+{^{n+1}}\,a^\dagger_-{^{n}}}{\sqrt{n!\,(n+1)!}}\,\, c^\dagger_{-1}\ket{0}\,,
\eal
with $n\geq 0$. In other words, on the odd sites $\text{L}_e=+1$ and thus $p_\phi=0$, while 
on the even ones $\text{L}_e=-1$ and $p_\phi=2$. In this representation, the wavefunction at time $t$ 
can be written as 
\bealn
\ket{\psi(t)} &= \sum_{\ell\geq 1}\,\psi_\ell(t)\ket{\ell}\,,
\eal
where the components $\psi_\ell(t)$ satisfy the equation of motion 
\bealn
i\,\dot{\psi}_\ell(t) &= \ep_\ell\,\psi_\ell(t) + t_{\ell-1}\,\psi_{\ell-1}(t) 
+ t_\ell\,\psi_{\ell+1}(t)\,,
\eal
with boundary conditions $\psi_\ell(0)=\delta_{\ell,1}$ and $\psi_{\ell=0}(t)=0$. Therefore, the unitary evolution of the initial state transforms into the propagation of a particle on the Lanczos chain that starts on the first site at $t=0$. It follows that 
\bealn
p_\phi(t) &= \bra{ \psi(t)}p_\phi\ket{\psi(t)}  = 2\sum_{n\geq 1}\, \big|\psi_{2n}(t)\big|^2\,.
\eal
In Fig.~\ref{FigureToy1} we plot the angular velocity $\dot{\phi}(t) = p_\phi(t) / I$. We observe that, since the dynamics does not account for the anticipated de-excitation of the electron, $\dot{\phi}(t)$ continues to oscillate between zero and a finite positive value.
However, if we considered the finite lifetime of the electronic excitation, and that were substantially longer than the time scale $1/\omega_0$, around $0.1~\text{ps}$ for AlF$_3$, we would expect a behavior similar to the time average of $\dot{\phi}(t)$ shown in Fig.~\ref{FigureToy1}, which converges to a finite value. We emphasize that this qualitative behavior holds irrespective of the harmonic approximation and the linear Jahn-Teller coupling. Indeed, the absorption of the circularly polarized light implies that a finite angular momentum is initially supplied to the electrons. This angular momentum is subsequently transferred to the molecular vibrations in the form of a pseudo angular angular momentum, see \eqn{pseudo angular momentum triangle} with the 1/2 eventually replaced by 1 if the second order Jahn-Teller coupling is dominant. However, since the molecular vibration does not carry any physical 
angular momentum, this process must inevitably be accompanied by a transfer of physical 
angular momentum to the molecule, which begins rotating.          \\

\begin{figure}[t]
  \centerline{\includegraphics[width=0.5\textwidth]{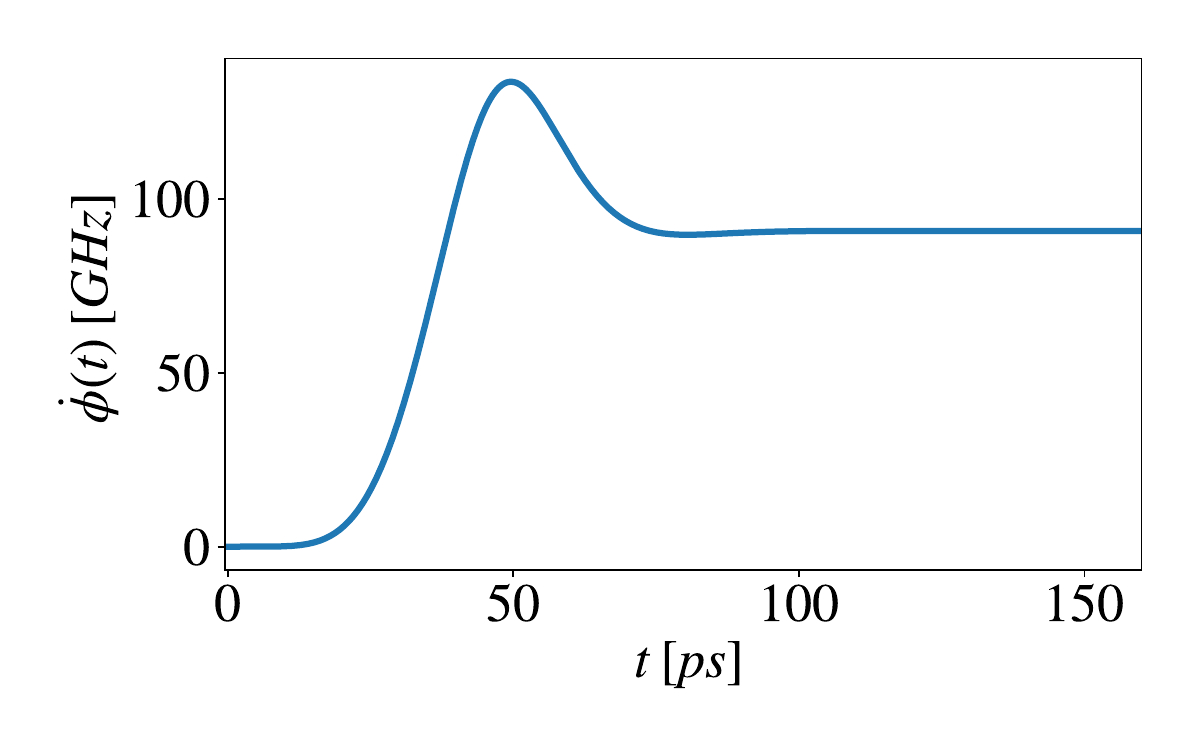}}
  \vspace{-0.2cm}
  \caption{Time evolution of $\dot{\phi}(t)$ induced by a circularly polarized electric field pulse in resonance with the chiral mode 1 of Fig.~\ref{FigureToy1}. For times longer than the pulse duration, i.e., $t \gtrsim 100\,\text{ps}$, $\dot{\phi}(t)$ becomes constant. The frequency and moment of inertia are the same 
as in Fig.~\ref{FigureToy1}. We take $\tau=33~\text{ps}$, the electric field 
maximum amplitude $E_0=800~\text{kV}/\text{cm}$, and estimate 
$d_*=0.1~e\text{\AA}$.} 
  \label{phidot_pulse}
\end{figure}
\noindent
Now, we delve into another hypothetical experiment that is more pertinent to our objectives. We observe that, because the molecule is ionic, 
the mode 1 with $\ell=\pm 1$ carries a dipole moment $\bd{d}=d_*\,(q_x,q_y)$, 
see Fig.~\ref{NF3-modes}, with $d_*$ the dipole strength, and therefore couples directly to an electric field. Let us therefore 
consider a circularly polarized electric field in the reference frame of the molecule, 
\bealn
\bd{E}(t) &= E(t)\,\big(\cos(\omega t-\phi),\sin(\omega t-\phi)\big)\,,
\eal
with $E(t)$ an envelope function finite in a time interval $t\in[0,\tau]$. With all MOs either fully occupied or empty, the Hamiltonian \eqn{H total triangle bis}
becomes
\beal
H(t) &= \fract{\;p_\phi^2\;}{2I} + \omega_0\,\Big(n_++n_-+1\Big)\\
&\quad -i\,\fract{d_*\,E(t)}{2}\,\bigg\{
\esp{i(\omega t-\phi)}\,\big(a^\dagger_+ + a^\dagga_-\big)\\
&\qquad\qquad\qquad\qquad
-
\esp{-i(\omega t-\phi)}\,\big(a^\dagga_+ + a^\dagger_-\big)
\bigg\}
\,,
\label{H total triangle tris}
\eal
and admits as conserved quantity the pseudo angular momentum
\beal
\text{J} &= p_\phi + n_+ - n_-\,.
\label{pseudo triangle}
\eal
We apply the time independent unitary transformation
\beal
U = \esp{-i\phi(n_+-n_-)}\;,
\label{unitary triangle}
\eal
after which $H(t) \to H'(t)=U^\dagger\,H(t)\,U$, where 
\beal
H'(t) &= \fract{1}{2I}\,\big(p_\phi-n_++n_-\big)^2 \\
&\qquad + \omega_0\,\Big(n_++n_-+1\Big)\\
&\quad -i\,\fract{d_*\,E(t)}{2}\,\bigg\{
\esp{i\omega t}\,\big(a^\dagger_+ + a^\dagga_-\big)\\
&\qquad\qquad\qquad\qquad
-
\esp{-i\omega t}\,\big(a^\dagga_+ + a^\dagger_-\big)
\bigg\}
\,.
\label{H total triangle quater}
\eal
We remark that $p_\phi$ is in fact the conserved pseudo angular momentum \eqn{pseudo triangle} after the unitary transformation. 
At $t<0$, before the field is turned on, we assume that the molecule has no excited boson 
and has $p_\phi=0$, which remains zero along the subsequent time evolution. If $\ket{\psi(t)}$ is the time-dependent wavefunction, solution of the 
Schr{\oe}dinger equation  $i\,\ket{\dot{\psi}(t)} = H'(t)\ket{\psi(t)}$, then
\bealn
\dot{\phi}(t) &= -i\,\bra{\psi(t)}\Big[\phi,H'(t)\Big]\ket{\psi(t)}\\
&= -\fract{1}{I}\, \bra{\psi(t)}n_+-n_-\ket{\psi(t)}\,,
\eal
which we show in Fig.~\ref{phidot_pulse}. We take an envelope function  
\bealn
 E(t)= E_0\,\left( \fract{t}{\tau} \right)^{2}  \; \exp\left\{1 -\left(\fract{t}{\tau}\right)^2 \right\} \,,
\eal
which describes an electric pulse that reaches a peak amplitude $E_0$ 
at $t=\tau$, and then decays exponentially fast. 
When the field is on, $0<t\gtrsim \tau$, $\dot{\phi }(t)$ increases until a maximum and then it converges to a lower but finite value for $t \gg \tau$, when the field is turned off 
and the Hamiltonian becomes again independent of $\phi$. Therefore, the electric field pulse is able to induce a permanent rotation of the entire molecule. 
We remark that the inclusion of high-energy modes and eventually anharmonic terms will cause $\dot{\phi}(t)$ to oscillate even for $t \gg \tau$, potentially leading to dephasing due to interference between different oscillation frequencies. Nevertheless, we anticipate that the long-time average of $\dot{\phi}(t)$ will remain finite as long as dissipation from the environment can be neglected.

\begin{figure}[t]
\centerline{\includegraphics[width=0.4\textwidth]{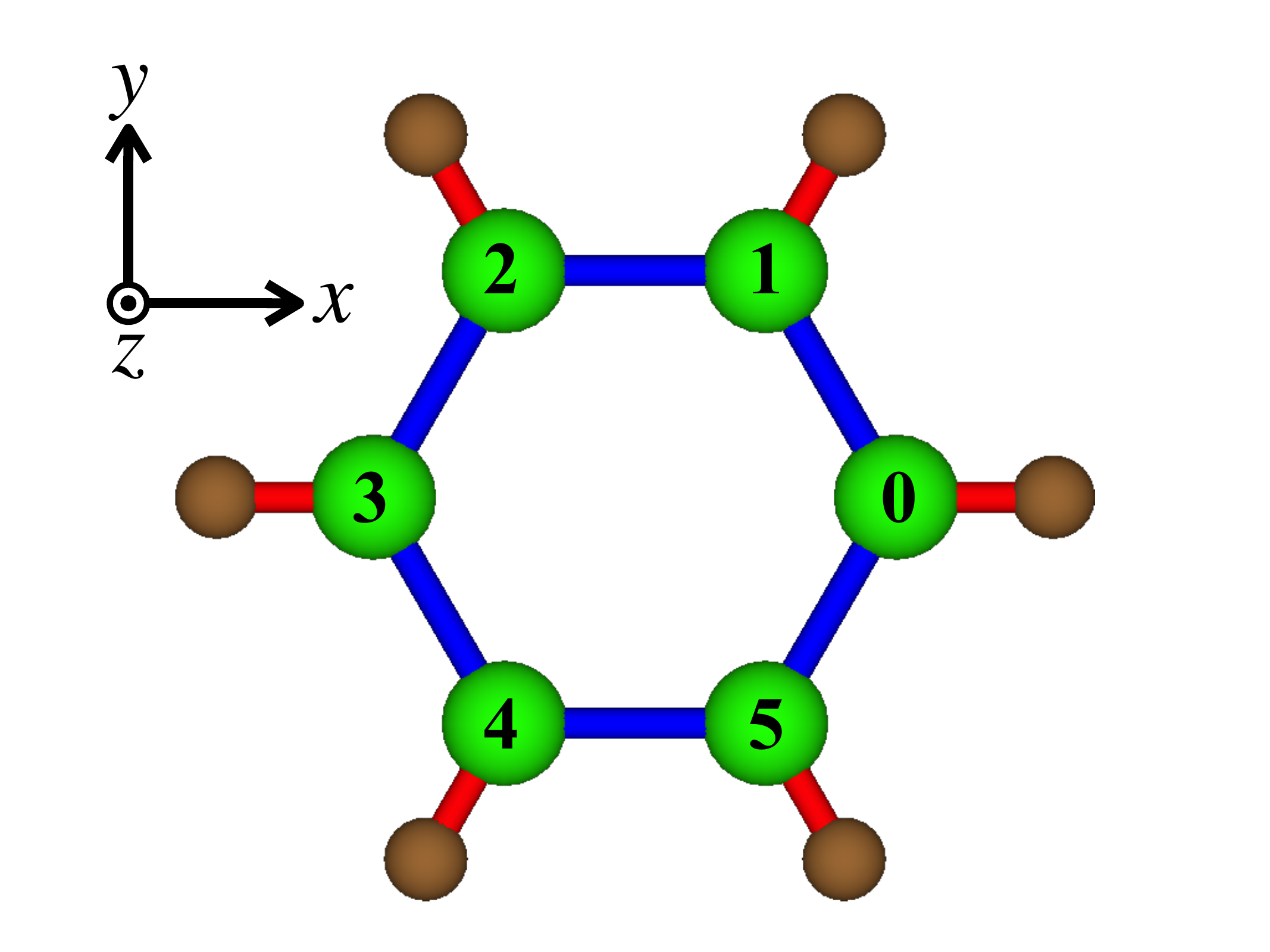}}
\caption{Toy planar molecule with $D_6$ symmetry. } 
\label{C6H6}
\end{figure}

\section{$D_6$ symmetric covalent molecule}
\label{$D_6$ symmetric covalent molecule}

The next example we consider is the planar molecule in Fig.~\ref{C6H6} that has $D_6$ symmetry. 
This molecule evidently reminds benzene, C$_6$H$_6$, with C atoms in green and H ones in brown. In what follows, we do use such analogy without pretending to give an accurate description of real benzene. \\
As before we use a reference frame in which the centre of the molecule is fixed, while we define as $\phi$ the angle that describes the rotation of the molecule around the $z$ axis, see Fig.~\ref{C6H6}. Furthermore, we project out the high-energy C-H breathing mode and neglect out-of-plane atomic displacements. Therefore, the point group of the molecule can be approximated with the 
two dimensional $D_6$ that contains the sixfold, $C_6$, threefold, $C_3$, and twofold, $C_2$, 
rotations around $z$, as well as twofold rotations around $x$, $C_2'$, and $y$, $C_2''$. 
Real benzene, due to its out-of-plane degrees of freedom, instead exhibits $D_{6h}$ point group symmetry, complemented by the $x$-$y$ mirror plane.
The character 
and product tables of the irreducible representations (irreps) of $D_6$ are shown in Table~\ref{Table D6}.
\begin{table}
\be\nonumber
\begin{array}{|c|c|c|c|c|c|c|}\hline
\bd{D_6} & E & C_6 & C_3 & C_2 & C'_2 & C''_2 \\ \hline
A_1 & 1 &  1 &  1 &  1 &  1 &  1 \\ \hline
A_2 & 1 &  1 &  1 &  1 & -1 & -1 \\ \hline
B_1 & 1 & -1 &  1 & -1 &  1 & -1 \\ \hline 
B_2 & 1 & -1 &  1 & -1 & -1 &  1 \\ \hline
E_1 & 2 &  1 & -1 & -2 &  0 &  0 \\ \hline
E_2 & 2 &  1 & -1 &  2 &  0 &  0 \\ \hline
\end{array}
\ee
\be\nonumber 
\begin{array}{|c||c|c|c|c|c|c|}\hline
~   & A_1 & A_2 & B_1 & B_2 & E_1 & E_2 \\ \hline\hline
A_1 & A_1 & A_2 & B_1 & B_2 & E_1 & E_2 \\ \hline
A_2 & A_2 & A_1 & B_2 & B_1 & E_1 & E_2 \\ \hline
B_1 & B_1 & B_2 & A_1 & A_2 & E_2 & E_1 \\ \hline
B_2 & B_2 & B_1 & A_2 & A_1 & E_2 & E_1 \\ \hline
E_1 & E_1 & E_1 & E_2 & E_2 & A_1\oplus[A_2]\oplus E_2 & B_1\otimes B_2\otimes E_1 \\ \hline
E_2 & E_2 & E_2 & E_1 & E_1 & B_1\otimes B_2\otimes E_1 & A_1\oplus[A_2]\oplus E_2 \\ \hline
\end{array}
\ee
\caption{Character table, top, and product table, bottom, of the wallpaper group $D_6$. The 
antisymmetric product of two identical irreducible representations is indicated by square brackets. }
\label{Table D6}
\end{table} 
Following the Appendix, the positions of the green atoms, $n=0,\dots,5$, in Fig.~\ref{C6H6} 
is parametrized through  
\bealn
\br_n &= \bR_n + \bx_n = \big(1+\alpha_n\big)\,\bR_n + \beta_n\,\bd{z}\wedge\bR_n\,,
\eal 
where $\bx_n$ is the displacement with respect to the equilibrium positions  
\bealn
\bR_n &= C_{\phi+(n-1)\pi/3}\big(\bR_0\big)\,, & \bR_0 &= (1,0)\,.
\eal
We make the change of variables \eqn{Fourier appx}
\beal
\begin{pmatrix}
\alpha_n\\
\beta_n
\end{pmatrix} 
&= \fract{1}{\sqrt{N}}\sum_{\ell}\,
\esp{i k_\ell n}\;\bq_\ell\,,
\label{Fourier hexagon}
\eal
now with 
\bealn
k_\ell &= \fract{2\pi}{6}\,\ell\,,& \ell &= -2,\dots,3\,. 
\eal
As discussed earlier, $\bq_0$ has only the first component finite, which corresponds to 
the $A_1$ breathing mode to which we associate the coordinate $q_{A_1}$. The displacement $\bq_{+1}=\bq_{-1}^*$ transforms like the 
irrep $E_1$, while $\bq_{+2}=\bq_{-2}^*$ like the 
irrep $E_2$. Finally, the first component $q_{B_1}$ of $\bq_3$ 
transforms like $B_1$, while the second component $q_{B_2}$ 
like $B_2$, thus the notation. The atomic Hamiltonian \eqn{H DN} reads in this case
\beal
H_\text{at} &= \fract{\;\big(p_\phi- \text{L}_\text{vib}\big)^2\;}{2I}+ 
\fract{\;p_{A_1}^2}{2M} 
+ \fract{\;p_{B_1}^2}{2M} + \fract{\;p_{B_2}^2}{2M} 
\\
&\qquad + \fract{1}{2M}\,\sum_{\ell\pm 1,\pm 2}\,\bp_{\ell}\cdot\bp_{-\ell} + V\big(\{\br_n\}\big)\,,
\label{H atomic C6H6}
\eal
where the moment of inertia is 
\be
I = M\,\big(\sqrt{6\,}+q_{A_1}\big)^2\,,
\label{moment of inertia D6}
\ee
and the displacement contribution to the angular momentum has the expression 
\be
\text{L}_\text{vib} = q_{B_1}\,p_{B_2} - q_{B_2}\,p_{B_1} 
+ \sum_{\ell\pm 1,\pm 2}\,\bq_\ell\wedge\bp_{-\ell}\cdot\bd{z}\,.
\label{L vibrational D6}
\ee
We note that the combinations that appear in \eqn{L vibrational D6} are consistent 
with $\text{L}_\text{vib}$ transforming like $A_2$ and the product Table~\ref{Table D6}.

\subsection{Inter atomic potential in the harmonic approximation}
As in the previous case, we discuss the normal modes assuming the harmonic approximation 
for the inter atomic potential $V\big(\{\br_n\}\big)$. We mention that anharmonic effects in benzene are not significant except for the C-H stretching modes \cite{Maslen-JCP1992},
which we do not take into account anyway.\\
We therefore consider the blue bonds in Fig.~\ref{C6H6}, which we approximate as springs with stiffness $K$ and equilibrium length that we take as unit of length. 
The inter atomic potential thus reads 
\be
V\big(\{\br_n\}\big) = \fract{K}{8}\sum_n\left[\big(\alpha_{n+1}+\alpha_n\big)
+ \sqrt{3}\,\big(\beta_{n+1}-\beta_n\big)\right]^2.
\label{potential benzene 1}
\ee
In addition to \eqn{potential benzene 1}, we include a further harmonic potential $\delta V$ 
with spring constant $\gamma\,K$, $\gamma\ll 1$ \cite{Medina03062015}, which is minimum when the angle between three successive carbon atoms is $120^\circ$ and refers to the fact that the energy is lower when the carbon atoms are in the $sp^2$ configuration. Specifically, 
\beal
\delta V\big(\{\br_n\}\big) &= \fract{\gamma K}{8}
\sum_n\,\Big(\alpha_{n+1}-\sqrt{3}\,\beta_{n+1}+\alpha_{n-1}\\
& \qquad\qquad\qquad\quad
+\sqrt{3}\,\beta_{n-1} -2\alpha_n\Big)^2\,.
\label{potential benzene 2}
\eal
The normal modes of the dynamical matrix solve an eigenvalue equation 
similar to \eqn{eigenvalue eq triangle 2}
\beal
\lambda_\ell^2\,\bu_\ell &= 
\begin{pmatrix}
\varepsilon_\ell + A_\ell & i B_\ell\\
-i B_\ell & \varepsilon_\ell - A_\ell
\end{pmatrix}\,\bu_\ell\,,
\label{normal mode eq benzene}
\eal
where $\lambda_\ell^2$ are the eigenvalues in units of $K$, and 
\beal
\varepsilon_\ell &= \fract{1}{2}\big(2 - \cos k_\ell +3\gamma 
-2\gamma \cos k_\ell - \gamma \cos 2k_\ell\big)\,,\\
A_\ell &= \fract{1}{2}\big(-1 + 2\cos k_\ell - 2\gamma\cos k_\ell +2\gamma\cos 2k_\ell\big)\,,\\
B_\ell &= \fract{\sqrt{3}}{2}\big( \sin k_\ell 
+ 2\gamma\sin k_\ell-\gamma\sin 2k_\ell\big)\,.
\eal
Upon defining 
\bealn
\theta_\ell &= -\theta_{-\ell} = \fract{1}{2}\,\tan^{-1}\fract{B_\ell}{A_\ell}\;,
\eal
the eigenvalues of \eqn{normal mode eq benzene} read
\bealn
\lambda_{1\ell}^2 &= \varepsilon_\ell - \sqrt{A_\ell^2+B_\ell^2\;}\;,&
 \lambda_{2\ell}^2 &= \varepsilon_\ell + \sqrt{A_\ell^2+B_\ell^2\;}\;,
\eal
with the corresponding eigenmodes
\bealn
\bu_{1\ell} &=  \begin{pmatrix}
i\sin\theta_\ell\\
\cos\theta_\ell
\end{pmatrix}\,,&
\bu_{2\ell} &=\begin{pmatrix}
\cos\theta_\ell\\
i\sin\theta_\ell
\end{pmatrix}\,.
\eal
Since $\gamma\ll 1$, $\lambda^2_{1\ell}\ll \lambda^2_{2\ell}$ for any $\ell$. 
For $\ell=0$, we must consider only the mode 2 that corresponds to $q_{A_1}$ and has eigenvalue 
$\lambda^2_{20}=\lambda^2_{A_1}$.   
Conversely, for $\ell=3$ mode 1 corresponds to $q_{B_2}$ and mode 2 to $q_{B_1}$, with 
eigenvalues $\lambda^2_{13}=\lambda^2_{B_2}$ and $\lambda^2_{23}=\lambda^2_{B_1}$, respectively. 
Finally, 
for $\ell=\pm 1,\pm 2$ we introduce the normal mode coordinates $q_{1\ell}$ and 
$q_{2\ell}$, as well as the corresponding conjugate momenta $p_{1\ell}$ and 
$p_{2\ell}$, so that, as before, 
\be\nonumber
\bq_\ell = q_{1\ell}\,\bu_{1\ell} + q_{2\ell}\,\bu_{2\ell}\,,
\ee
and similarly for $\bp_\ell$. By symmetry, the modes at $\ell$ and $-\ell$ are degenerate.
The atomic Hamiltonian \eqn{H atomic C6H6} within the harmonic approximation thus reads
\beal
&H_\text{at} = \fract{1}{2I}\,\Big(p_\phi-\text{L}_\text{vib}\Big)^2 
+ \fract{\;p_{A_1}^2}{2M} + \fract{K\lambda_{A_1}^2}{2}\;q_{A_1}^2\\
&\qquad + \fract{\;p_{B_1}^2}{2M} + \fract{K\lambda_{B_1}^2}{2}\;q_{B_1}^2\\
&\qquad + \fract{\;p_{B_2}^2}{2M} + \fract{K\lambda_{B_2}^2}{2}\;q_{B_2}^2\\ 
&\quad + \sum_{a=1,2}\sum_{\ell=\pm 1,\pm 2}
\Bigg( \fract{1}{M}\,p_{a\ell}^\dagga\,p_{a\ell}^\dagger 
+ K\,\lambda_{a\ell}^2\,q_{a\ell}^\dagga\,q_{a\ell}^\dagger\Bigg)\,.
\eal
where the contribution of the normal modes to the total angular momentum is 
\beal
\text{L}_\text{vib} &= \Big(q_{B_1}\,p_{B_2} - q_{B_2}\,p_{B_1}\Big)\\
&\; +\sum_{\ell=1}^2\,\bigg\{
i\,\sin2\theta_\ell\,\big(q_{1\ell}\,p_{1-\ell} - q_{1-\ell}\,p_{1\ell}\\
&\qquad\qquad \qquad\qquad\quad
-q_{2\ell}\,p_{2-\ell} + q_{2-\ell}\,p_{2\ell}\big)
\\
&\qquad\qquad\;
 -\cos2\theta_\ell\,\Big(q_{1\ell}\,p_{2-\ell} + q_{1-\ell}\,p_{2\ell}\\
&\qquad\qquad \qquad\qquad\quad
-q_{2\ell}\,p_{1-\ell} - q_{2-\ell}\,p_{1\ell}\Big)\bigg\}
\,.
\label{L-vib-benzene}
\eal
We observe that the two combinations 
\beal
L^+_1 &= \cos\left(\fract{\pi}{4}-\theta_1\right)\, q_{1+1} \\
&\qquad -i\,\sin\left(\fract{\pi}{4}-\theta_1\right)\, q_{2+1}\,,\\
L^+_2 &= i\,\sin\left(\fract{\pi}{4}-\theta_1\right)\,q_{1-1}\\
&\qquad 
+\cos\left(\fract{\pi}{4}-\theta_1\right)\,q_{2-1} 
\,,
\label{rising}
\eal
act like rising operators for $\text{L}_\text{vib}$ in \eqn{L-vib-benzene}, thus the hermitian conjugates of \eqn{rising} as lowering ones. One can readily obtain analogous operators 
for $\ell=\pm 2$ and for the $B$ modes.

\subsection{Electron Hamiltonian}
\label{Electron Hamiltonian}

We assume that each green atom in Fig.~\ref{C6H6} hosts a $p_z$ orbital occupied by one electron. At fixed atomic positions, the 
electron Hamiltonian is simply a tight-binding on a hexagon with nearest, $t_1$, next nearest, $t_2$, and next to next nearest, $t_3$, hopping amplitudes, with $t_1>t_2>t_3>0$. Therefore, the 
single-particle eigenstates have eigenvalues 
\bealn
\ep_\ell &= -2\,\sum_{j=1}^3\,t_j\,\cos k_\ell j
\,,& \ell&=-2,\dots,3\,.
\eal
We assume that the hopping amplitudes are such that 
$\ep_3 > \ep_{\pm 2} > \ep_{\pm 1} > \ep_0$. 
The state with $\ell=0$ transforms like the $A_1$ irrep of $D_6$, that with $\ell=3$ like the $B_1$ irrep, 
while the two doublets with $\ell =\pm 1$ and $\ell=\pm 2$, respectively, like the two dimensional irreps $E_1\sim Y_{1,\pm 1}$ and $E_2\sim Y_{2,\pm 2}$. 

\subsubsection{Electronic moment of inertia and magnetic orbital moment}
\label{Electronic moment of inertia}

We suppose that the molecule rotates with constant angular 
velocity $\omega$, thus the green atom equilibrium positions are time-dependent $\bR_n\to\bR_n(t)$, $n=0,\dots,5$. The single-electron Hamiltonian in first quantization is 
therefore 
\bealn
H(t) &=  \fract{\bp^2}{2m} + \sum_{n=0}^5\,U\big(\br-\bR_n(t)\big)\,,
\eal
with $m$ the electron mass, and the Schr\"{o}dinger equation reads, accordingly,
\bealn
i\,\fract{d\psi(\br,t)}{dt} = H(t)\,\psi(\br,t)\,.
\eal
We use the standard trick and make a change of variable using the reference frame in which the atoms are at rest, thus
\bealn
\br &\to \br(t) = \cos\omega t\;\br + \sin\omega t\;\br\times\bd{z}\,,
\eal
and assume $\psi(\br,t)\to\phi(\br(t),t)$ so that 
\bealn
i\,\fract{d\phi(\br(r),t)}{dt} &= 
i\,\fract{\partial\phi(\br(r),t)}{\partial t} \\
&\qquad 
-\omega\,\br(t)\times\bd{z}\cdot\bp\,\phi(\br(r),t)\,.
\eal
In this reference frame $U(\br-\bR_n(t))\to U(\br(t)-\bR_n)$. 
Since only the explicit time-derivative appears, we can replace $\br(t)\to\br$ 
and get the effective Schr\"{o}dinger equation 
\beal
i\,\fract{\partial\phi(\br,t)}{\partial t} &= 
\Big\{H 
+\omega\,\br\times\bd{z}\cdot\bp\Big\}\,\phi(\br,t) \\
&= H(\omega)\,\phi(\br,t)\,,
\label{H*(omega)}
\eal 
now with time-independent Hamiltonian. The solution is simply obtained: 
if $\phi_a(\br)$ is any eigenstate of $H(\omega)$ with eigenvalue $\ep_a(\omega)$, then 
$\phi_a(\br,t)=\esp{-i\ep_a(\omega) t}\;\phi_a(\br)$. As a matter of fact, $H(\omega)$ looks 
like the Hamiltonian in presence of a fictitious magnetic field  
generated by a vector 
potential 
\bealn
\bd{A}(\br) &= \fract{\omega m c}{e}\, \br\times \bd{z} = \fract{\omega m c}{e}\,\big(y,-x\big)\,.
\eal 
In reality, 
\bealn
&\fract{\bp^2}{2m}  +\fract{e}{mc}\,\bd{A}(\br)\cdot\bp \\
&= \fract{1}{2m}\bigg(\bp  +\fract{e}{c}\,\bd{A}(\br)\bigg)^2 - \fract{m\omega^2}{2}\,r^2\\
&\simeq \fract{1}{2m}\bigg(\bp  +\fract{e}{c}\,\bd{A}(\br)\bigg)^2 - \fract{m\omega^2}{2}\,,
\eal
since the electron coordinates are localized around the atomic positions, all of which 
are at unit distance from the origin. \\
Assuming the Peierls approximation and, for simplicity, just the nearest neighbor hopping $t_1$, the fictitious magnetic field adds a phase 
\bealn
\Phi &= \fract{\sqrt{3}}{2}\,\omega\, m\,,
\eal
so that the eigenstates of $H(\omega)$ are the same $\psi_\ell(\br)$ of the tight-binding Hamiltonian, but now have eigenvalues 
\bealn
\varepsilon_\ell(\omega) &= -2\,t_1\,\cos \big(k_\ell + \Phi\big) 
- \fract{m\omega^2}{2} \\
&= \ep_\ell(\omega)- \fract{m\omega^2}{2}\;.
\eal
I note that, since $\bd{A}(\br)\propto \omega$,  
\bealn
\fract{\partial H(\omega)}{\partial \omega} &= \fract{e}{\omega mc}\,\bd{A}(\br)\cdot\bp\,,
\eal
so that 
\bealn
&\int d\br\,\psi_\ell(\br)^*\,\bigg(\fract{e}{mc}\,\bd{A}(\br)\cdot\bp\bigg)
\psi_\ell(\br) \\
&= \omega\,\int d\br\,\psi_\ell(\br)^*\;\fract{\partial H(\omega)}{\partial \omega}\;
\psi_\ell(\br)\\
& = \omega\,\fract{\partial\varepsilon_\ell(\omega)}{\partial\omega}
=  \Phi\;\fract{\partial\ep_\ell(\omega)}{\partial\Phi} 
-m\omega^2\,,
\eal
because also $\Phi\propto \omega$. 
The actual value $E_\ell(\omega)$ of the energy of the state $\psi_\ell(r,t)=\esp{-i\ep_\ell(\omega) t}\;\phi_\ell(\br(t))$ expanded at second order in $\omega$
is
\beal
&E_\ell(\omega) = i\,\int d\br\,\psi_\ell(r,t)^*\;
\fract{d\psi_\ell(r,t)}{dt}\\
&\quad = \int d\br\,\psi_\ell(\br)^*\,\bigg(\varepsilon_\ell(\omega) -\fract{e}{mc}\,\bd{A}(\br)\cdot\bp\bigg)
\psi_\ell(\br)\\
&\qquad = \ep_\ell(\omega) 
- \fract{m\omega^2}{2}-\Phi\;\fract{\partial\ep_\ell(\omega)}{\partial\Phi} 
+m\omega^2\\
&\qquad \simeq \ep_\ell + \fract{m\omega^2}{2} + \fract{3}{8}\,\omega^2\,m^2\,\ep_\ell
\,,
\eal
which implies that the moment of inertia of an electron in the wavefunction $\psi_\ell(\br)$ is
\bealn
I_\ell &= m\,\left( 1 + \fract{3}{4}\,m\,\ep_\ell\right)\,.
\eal
We note that $\sum_\sigma\,\sum_\ell\,I_\ell = 12m$, so that, if all levels are occupied, the electronic contribution to the moment of inertia is the expected one. 
On the contrary, in the case with 6 electrons that we consider, the moment of inertia 
\bealn
I_e &= \sum_\sigma\,\sum_{\ell=-1}^1\,I_\ell = 6m - 3t_1\,m^2\,,
\eal
is reduced with respect to the bare value $6m$, a kind of diamagnetic response to  
angular rotations.\\
We can repeat the calculations above assuming that the molecule is pierced by a real magnetic flux and readily find that each electron state carries a magnetic orbital moment $\text{L}_\ell$ given by
\beal
\text{L}_\ell &= \mu_B\,\sqrt{3\,}\,m\,t_1\,\sin\fract{\pi\ell}{3}\;.
\eal

\subsection{Coupling to the normal modes}
\label{Coupling to the normal modes}

We assume that the hopping between sites $n$ and $n+m$ is a decaying function of their distance 
$|\br_{n+m}-\br_n|$. It follows that, for small displacements $\bx_n$ and $\bx_{n+m}$, $m=1,2,3$, 
\beal
t_m\big(|\br_{n+m}-\br_n|\big) &\simeq t_m\Big\{1 - g_m c_m\,\big(\alpha_{n+m}+\alpha_n\big) \\
&\qquad\qquad\;
-g_m s_m\,\big(\beta_{n+1} -\beta_{n}\big)\Big\}\\
&\equiv t_m -\delta t_m\,,
\eal
where $g_m$ is the coupling constant and 
\bealn
c_m &= \cos\fract{\pi(3-m)}{6}\;,&
s_m &= \sin\fract{\pi(3-m)}{6}\;.
\eal
It follows that the coupling between the electrons and the normal modes is 
\beal
V_\text{el-vib} &= \sum_{m=1}^3\,\delta t_m\,\sum_{n\sigma}\, \big(c^\dagger_{n+m\sigma}\,c^\dagga_{n\sigma}+c^\dagger_{n\sigma}\,c^\dagga_{n+m\sigma}
\big)\\
&= \fract{1}{\sqrt{6}}\sum_m\,g_m\,t_m\,\sum_\sigma\,\sum_{\ell_1\ell_2}\,
c^\dagger_{\ell_1\sigma}\,c^\dagga_{\ell_2\sigma}\\
&\qquad\qquad \qquad\qquad
\Big(\esp{-i m k_{\ell_1}} + \esp{imk_{\ell_2}}\Big)\\
&\qquad\qquad 
\bigg\{ c_m\,a_{\ell_1-\ell_2}\,\Big(\esp{imk_{\ell_1-\ell_2}} + 1\Big)\\
&\qquad\qquad \quad 
+ s_m\,b_{\ell_1-\ell_2}\,\Big(\esp{imk_{\ell_1-\ell_2}} - 1\Big)\bigg\}\,,
\label{el-ph coupling benzene}
\eal
where $\ell_2+\ell_3$ or $\ell-\ell_2$ are defined modulo 6 with integers between 
-2 and 3. It is worth emphasizing that the momentum conservation in \eqn{el-ph coupling benzene} 
simply reflects the invariance of the Hamiltonian with respect to the $D_3$ point group.\\
We note that the coupling to the normal modes vanishes identically whenever 
$\esp{-imk_{\ell_1}} + \esp{imk_{\ell_2}} = 0$, namely, when 
\bealn
m\,\big(\ell_1+\ell_2\big) &= 3(1+2n)\,,& m&=1,2,3\,.
\eal
In particular, the normal mode assisted tunneling between the electronic states with $\ell=\pm 1$ 
and those with $\ell=\pm 2$ only occurs via the next-nearest neighbor hopping $t_2$, an observation 
that we use later. 

\subsection{Coupling to a circularly polarized electromagnetic field} 
\label{Coupling to a circularly polarized electromagnetic field}

We assume that a circularly polarized electromagnetic field (EMF) is applied to the molecule, with 
\bealn
\bd{E}(t) &= E\,\big(\cos\omega t,\sin\omega t\big)\,,
\eal
and thus vector potential 
\bealn
\bd{A}(t) &= \fract{cE}{\omega}\,\big(-\sin\omega t,\cos\omega t\big)\,.
\eal
In the reference frame of the molecule, which is rotated by $\phi$, 
\bealn
\bd{A}(t,\phi) &= \fract{cE}{\omega}\,\Big(\sin(\phi-\omega t),\cos(\phi-\omega t)\Big)\,.
\eal
The bonds that connect atom $n$ with $n+m$ are the vectors $\bR_{n,m}$, $m=1,2,3$, given by
\beal
&\bd{R}_{n,m} =\bR_{n+m} - \bR_n \\
&\;= \bigg(
\cos\fract{(2n+3+m)\pi}{6}\,,\,\sin\fract{(2n+3+m)\pi}{6}\bigg)\,.
\eal
It follows that the Peierls phase acquired by the hopping term $c^\dagger_{n+m\sigma}\,c^\dagga_{n\sigma}$ is 
\bealn
\phi_{n,m} &= \fract{e}{c}\,\bd{A}(t,\phi)\cdot\bd{R}_{n,m} \\
&= \fract{eE}{\omega}\,\sin\Big((2n+3+m)\pi/6-\phi -\omega t\Big)\\
&= -i\,\fract{eE}{2\omega}\,\Big(
\esp{i\frac{\pi}{6}(3+m)}\;\esp{i(\phi-\omega t)}\;\esp{i n\pi/3} - c.c.\Big)\\
&\equiv Q_m\,\esp{i(\phi-\omega t)}\;\esp{i n\pi/3} + c.c.
\,,
\eal
and behaves like the sum of a Bloch wave with momentum $k_1$ and one with $k_{-1}$. 
The first order correction in the EMF to the hopping Hamiltonian thus reads
\beal
\delta T &= -i\, \esp{i(\phi-\omega t)}\,\sum_m\,Q_m\,t_m\,\sum_\sigma\,\sum_{\ell_1\ell_2\ell_3}\,
c^\dagger_{\ell_1\sigma}\,c^\dagga_{\ell_2\sigma} \\
&\qquad 
 \bigg(\esp{-ik_{\ell_1}m} - \esp{ik_{\ell_2}m}\bigg) \Bigg\{\delta_{\ell_1,\ell_2+1}\\
&\qquad\quad   - \fract{g_m}{\sqrt{6}}\,c_m\,a_{\ell_1-\ell_2-1}\,
\bigg(\esp{ik_{\ell_1-\ell_2-1}m} + 1\bigg)\\
&\qquad\quad - \fract{g_m}{\sqrt{6}}\,s_m\,b_{\ell_1-\ell_2-1}\,
\bigg(\esp{ik_{\ell_1-\ell_2-1}m} - 1\bigg)\Bigg\}\\
& + H.c.\,.
\label{delta T}
\eal
As expected, the circularly polarized light can induce on its own electronic excitations with $\Delta\ell=\pm 1$, or with different $\Delta\ell$ in cooperation with the molecular vibrations. 
We also note from \eqn{delta T} that, if the electronic configuration is prepared in a finite current state, i.e., with  
\bealn
J &= \sum_m\,t_m\,\sum_{\ell\sigma}\, \sin k_\ell m\; \langle\,c^\dagger_{\ell\sigma}\,c^\dagga_{\ell\sigma}\,\rangle \not=0\,,
\eal 
then the light couples directly to the normal modes with $\ell=\pm 1$, despite the molecule is non-polar and thus the vibrations are not charged. This is not the sole such possibility, as we are going to show. 

\subsection{Hypothetical experiment}
\label{Hypothetical experiment benzene}
\begin{figure}[t]
 \centerline{\includegraphics[width=0.5\textwidth]{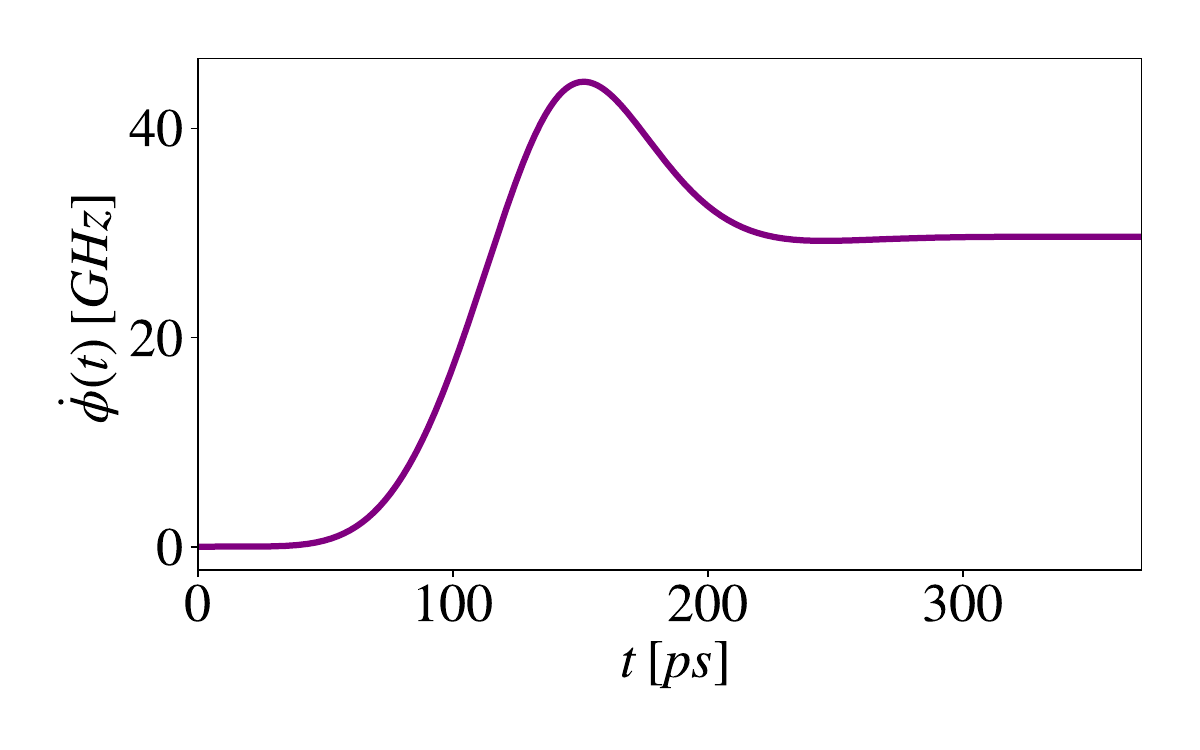}}
  \vspace{-0.2cm}
  \caption{Time evolution of $\dot{\phi}(t)$ in the benzene-like molecule induced by a circularly polarized electric field pulse in resonance with the $E_{1u}$ C-H bending mode. The qualitative behavior is the same as the one shown for the $D_3$ molecule in Fig.~\ref{phidot_pulse}. The C-C bond length is taken as 
  $1.413\,\text{\AA}$ \cite{Medina03062015}, 
  while the frequency of the $E_{1u}$ phonon as 
  $\omega_0= 1048 \,\text{cm}^{-1}$ \cite{fononi_benzene}. The dipole associated to the phonon is $d_{\ast}= 9.5 \cdot 10^{-4}\, e \text{\AA}$ \cite{dipolo_benzene} and the moment of inertia in units of $\hbar^{2}/\omega_0$ is $I\simeq 123 $. We take $\tau= 101  \, \text{ps} $ and the electric field maximum amplitude  $E_0=10~\text{MV}/\text{cm}$.   }
  \label{phidot_pulse_benzene}
\end{figure}

We assume that the molecule contains six electrons and therefore the electronic ground state has occupied 
$\ell=0,\pm 1$ levels. If the EMF frequency $\omega$ and the energy of the normal mode 1 with 
$\ell=\pm 1$ are close to each other and much smaller than $\ep_{\ell=\pm2}-\ep_{\ell=\pm1}\equiv\Delta$, a coupling between the electromagnetic field and the normal mode is generated 
by second order perturbation theory \cite{Rice&Choi}. For simplicity, we only consider in \eqn{delta T} the field induced excitations 
from the occupied $\ell=\pm 1$ to the empty $\ell=\pm 2$ 
without involvement of the molecular vibrations and limited to the dominant nearest neighbor hopping. 
In this case,
\beal
\delta T &\simeq -\fract{eE t_1}{\omega}\sum_\sigma
\Big\{\esp{i(\phi-\omega t)}\;
c^\dagger_{2\sigma}\,c^\dagga_{1\sigma} \\
&\qquad \qquad\qquad\quad
+\esp{-i(\phi-\omega t)}\;
c^\dagger_{-2\sigma}\,c^\dagga_{-1\sigma}\Big\}+ H.c.\\
&\equiv \delta T^+ + \delta T^-\,.
\label{delta T bis}
\eal
Concerning the coupling between electrons and normal modes, we just focus on the mode with $\ell=\pm 1$ 
that is also able to transfer one electron from $\ell=\pm 1$ to $\ell=\pm 2$. 
As previously discussed, the coupling term \eqn{el-ph coupling benzene} vanishes 
for nearest neighbor hopping. We therefore consider the next-nearest neighbor one, in which case
\eqn{el-ph coupling benzene} reads, through \eqn{rising},  
\beal
V_\text{el-vib}&\simeq -i\,g_2\,t_2\sum_\sigma\Big\{
c^\dagger_{2\sigma}\,c^\dagga_{1\sigma}\;L^+_1 + c^\dagger_{-2\sigma}\,c^\dagga_{-1\sigma}\;
L^-_1\Big\}\\
&\qquad\qquad  + H.c. \equiv V_\text{el-vib}^+ + V_\text{el-vib}^-\,.
\label{el-ph coupling benzene bis}
\eal
It follows that, if $\ket{n}$ are the electronic eigenstates with eigenvalues $E_n$, with $\ket{0}$ the ground state, then, by second order perturbation theory, 
\beal
&V_\text{EMF-vib} = \sum_{n>0}\,\Bigg\{
\fract{\bra{0}\delta T^-\ket{n}\bra{n}V_\text{el-vib}^+\ket{0}}
{E_n - E_0} \\
&\qquad\qquad\qquad +
\fract{\bra{0}V_\text{el-vib}^-\ket{n}\bra{n}\delta T^+\ket{0}}
{E_n - E_0}\Bigg\}\\
&\quad = 2i\;\fract{eE}{\omega}\;\fract{g_2\,t_1\,t_2}{\Delta}\,
\bigg\{\esp{-i(\phi-\omega t)}\;L^+_1 +H.c.\bigg\}\,.
\label{Michael Rice}
\eal
In other words, the virtual excitation of a particle-hole pair with $\Delta\ell=\pm 1$ mediates 
a coupling between the EMF and the chargeless normal modes, rendering the latter observable in the infrared spectrum \cite{Rice&Choi}.  
We also note that \eqn{Michael Rice} has a conserved pseudo angular momentum 
\beal
\text{J} &= p_\phi + \text{L}_\text{vib} = \text{L} + \text{L}_\text{vib}\,.
\label{pseudo benzene}
\eal
Therefore, the EMF is able not only to excite the normal modes with $\ell=\pm 1$ but also 
to change the total angular momentum $\text{L}$ of the molecule. \\
Since $\gamma\ll 1$, the angle $\theta_1\simeq \pi/4$, thus $L^+_1\simeq q_{1+1}$ and
\beal
V_\text{EMF-vib} &\simeq 2i\;\fract{eE}{\omega}\;\fract{g_2\,t_1\,t_2}{\Delta}\\
&\qquad \bigg\{\esp{-i(\phi-\omega t)}\;q_{1+1}
-\esp{i(\phi-\omega t)}\;q_{1-1}\bigg\}\,.
\label{Michael Rice bis}
\eal
We define dimensionless normal mode coordinates through  
\bealn
q_{1\pm 1} &\to \sqrt{K_1\;}\;q_{1\pm 1}\,,& p_{1\pm 1} &\to \fract{1}{\;\sqrt{K_1\,}\;}\;
p_{1\pm 1}\,,
\eal
with $K_1=1/M\omega_0$ and $\omega_0 = \lambda_{1\pm 1}\,\sqrt{K/M}$ the normal mode energy. 
The $a=1$ and $\ell=\pm 1$ contribution to $\text{L}_\text{vib}$ is therefore 
\beal
\text{L}_\text{vib} &=  i\,\Big(q_{1+1}\,p_{1-1} - q_{1-1}\,p_{1+1}\Big)
\,.
\label{L2-benzene-1}
\eal 
We further define 
\bealn
q_{1\,+1} &=  \fract{1}{\sqrt{2}}\big(a^\dagga_{-}
+a^\dagger_{+}\big)\,,&
p_{1\,-1} &= 
\fract{i}{\sqrt{2}}\big(a^\dagger_{-}-a^\dagga_{+}
\big)
\,,
\eal
so that 
\beal
\text{L}_\text{vib} &=  a^\dagger_{+}\,a^\dagga_{+} - a^\dagger_{-}\,a^\dagga_{-} = 
n_{+}-n_{-}\,. 
\label{L2-benzene-1 bis}
\eal  
The simplified Hamiltonian $H(t)$ in presence of the EMF reads therefore 
\beal
H(t) &= \fract{1}{2I}\,\Big(p_\phi-n_{+}+n_{-}\Big)^2 \\
&\qquad +\omega_0\,\Big(n_+ + n_- + 1\Big)\\
&\qquad + i\,E(t)\,\gamma\,\Big\{\esp{-i(\phi-\omega t)}\;\big(a^\dagga_- + a^\dagger_+\big) \\
&\qquad\qquad\qquad \qquad
-\esp{i(\phi-\omega t)}\;\big(a^\dagga_+ + a^\dagger_-\big)\Big\}\,,
\label{H(t)}
\eal
where $\gamma$ is a constant and $E(t)$ an envelope function that grows from zero at $t>0$ and remains finite for a finite time interval $\tau$. We observe that \eqn{H(t)} resembles \eqn{H total triangle tris}, except that the normal modes now carry an angular momentum that enters 
explicitly in \eqn{H(t)}. That would make the molecule rotate even if $\phi$ did not appeared explicitly. We plot the time evolution of $\dot{\phi}$ in Fig.~\ref{phidot_pulse_benzene}. Unsurprisingly, the dynamical behavior remains qualitatively the same as that 
in Fig.~\ref{phidot_pulse} of Sec.~\ref
{Hypothetical experiment triangle}.

\section{Conclusions}
We have shown how circularly polarized light can induce angular rotation of molecules that possess coupled doubly degenerate electronic orbitals and (chiral) normal modes that realize an $e\times E$ Jahn-Teller effect. This coupling is invariant under $U(1)$ rotations with generator a pseudo angular momentum distinct from the physical one. In particular, we have considered two exemplary cases: a $D_3$ symmetric ionic molecule and a $D_6$ symmetric covalent one, inspired, respectively, by metal trifluorides and by benzene. The Jahn-Teller effect originates from the dependence of the crystal field on the ion displacements within the ionic molecule, and, in the covalent case, of the electron covalent bonding on the atomic displacements, thereby providing a comprehensive representation of all possible physical realizations. \\
Although our results are specific to two molecular toy models, certain aspects might be relevant also to bulk systems. For instance, the interaction between circularly polarized light and the chiral normal modes of bulk crystalline samples can be direct in ionic materials or mediated by virtual particle-hole 
excitations in covalent systems. In either case, this coupling  
depends on the relative angle $\phi$ between the sample reference frame and the laboratory frame. Consequently, the light can drive the sample into rotation, inducing a finite angular velocity 
$\dot{\phi}$.\\
Furthermore, if the chiral modes independently contribute to the physical angular momentum, as observed in the benzene molecule, their coherent excitation, regardless of its realization \cite{Hamada-PRL2018,Zhong-PRB2023,zhang-ArXiv2024,Zhang-NanoLett2024}, might yield comparable effects to circularly polarized light. 
 
\begin{acknowledgments}
 The authors are very grateful to Antimo Marrazzo for stimulating discussions that inspired the work.
\end{acknowledgments}
\appendix*

\section{Derivation of the Hamiltonian}
\label{appendix}
In this Appendix, we derive the expression for the atomic Hamiltonian of a planar molecule with a regular $N$-sided polygon shape and equivalent atoms of mass $M$ at its vertices.
As in the main text, we disregard out-of-plane displacements. Consequently, the point group of the molecule is the two-dimensional $D_N$. We also neglect the motion of the center of mass, which coincides with the center of the polygon, but we account for the molecule's rotations around this point. Our objective is to separate the global molecule rotation from all internal degrees of freedom, which is straightforward in this regular geometry.  
\\
The atoms at the vertices are at position $\br_n$, $n=0,\dots,N-1$, which we parametrize as 
\beal
\br_n = \bR_n + \bx_n = \big(1+\alpha_n\big)\,\bR_n + \beta_n\,\bd{z}\wedge\bR_n\,,
\label{r-n DN}
\eal
where $\bR_n$ are the equilibrium positions, $\bd{z}$ the unit vector perpendicular to the molecule plane, and $\bx_n$ the displacements. By symmetry, 
\bealn
\bR_n &= C_{n 2\pi/N}\big(\bR_0\big)\,,& n&=0,\dots,N-1\,,
\eal
where $C_\theta$ is a rotation by an angle $\theta$ around the centre of the molecule. 
Hereafter, we set $|\bR_0|=1$ our unit of length.\\
We denote by $\phi$ the dynamical rotation angle of the molecule with respect to a fixed reference frame. Consistently, we must exclude the displacements that also 
correspond to a global molecule rotation. With the parametrization \eqn{r-n DN}, this 
amounts to impose that 
\be
\sum_n\,\beta_n = \sum_n\,\bx_n\cdot\bd{z}\wedge\bR_n 
= \sum_n\,\br_n\cdot\bd{z}\wedge\bR_n\equiv  0\,,
\label{constraint}
\ee
which implies that also the time derivatives vanish. To enforce \eqn{constraint}, we 
make the unitary transformation 
\beal
\begin{pmatrix}
\alpha_n\\
\beta_n
\end{pmatrix} &= \fract{1}{\sqrt{N}}\sum_{\ell}\,
\esp{i k_\ell n}\; \begin{pmatrix}
a_\ell\\
b_\ell
\end{pmatrix}\\
&= \fract{1}{\sqrt{N}}\sum_{\ell}\,
\esp{i k_\ell n}\;\bq_\ell\,,
\label{Fourier appx}
\eal
where $k_\ell = 2\pi\ell/N$, $\ell= -\lceil N/2\rceil +1,\dots,\lfloor N/2\rfloor$ 
are the wave vectors corresponding to an $N$-site 
chain with periodic boundary conditions, and $\bq_\ell^\dagga=\bq_{-\ell}^*$. 
In this representation, the constraint \eqn{constraint} simply implies that 
$b_0\equiv 0$. Therefore, $\bq_0$ has only one component $a_0$. Another advantage of \eqn{Fourier appx} is that $\bq_\ell$ 
transform automatically as the irreducible representations of the point group. 
For instance, $a_0$ represents the $A_1$ breathing mode, and so we define $a_0\coloneq q_{A_1}$. \\  
With respect to the fixed reference frame, the atomic positions read
\bealn
\br_n(\phi) &= C_\phi\big(\br_n\big) = \cos\phi\,\br_n + \sin\phi\,\bd{z}\wedge\br_n\,,
\eal
thus 
\be
\dot{\br}_n(\phi) = C_\phi\big(\dot{\br}_n + \dot{\phi}\,\bd{z}\wedge\br_n\big)
= C_\phi\big(\dot{\bx}_n + \dot{\phi}\,\bd{z}\wedge\br_n\big)
\,.
\ee
Therefore, taking the constraint \eqn{constraint} into account and using \eqn{Fourier appx}, the angular momentum $L$ has the expression 
\beal
\text{L} &= M\,\sum_n\,\br_n(\phi)\wedge \dot{\br}_n(\phi)\cdot\bd{z}\\
&= M\,\sum_n\,\br_n\wedge\big(\dot{\br}_n + \dot{\phi}\,\bd{z}\wedge\br_n\big)\cdot\bd{z}\\
&= \dot{\phi}\,M\,\sum_n\,\text{r}_n^2 + 
M\,\sum_n\,\bx_n\wedge\dot{\bx}_n\cdot\bd{z}\\
&\coloneq \dot{\phi}\,I_0 + M\,\sum_{\ell\not=0}\,\bq_\ell\wedge\dot{\bq}_{-\ell}\cdot\bd{z}
\coloneq \dot{\phi}\,I_0 + J\,,
\label{L DN}
\eal
where we remark that $\ell=N/2\equiv -N/2$ for even $N$, and we can write
\bealn
J &=  M\,\sum_{\ell\not=0}\,\bq_\ell\wedge\dot{\bq}_{-\ell}\cdot\bd{z}
= M\,\sum_{\ell\not=0}\,\dot{\bq}_{\ell}\cdot\bd{z} \wedge \bq_{-\ell}\,,\\
I_0 &= M\,\big(\sqrt{N}+q_{A_1}\big)^2 + M\,\,\sum_{\ell\not=0}\,
\bq_\ell\cdot\bq_{-\ell}\\
&\coloneq I + M\,\,\sum_{\ell\not=0}\,
\bq_\ell\cdot\bq_{-\ell} \coloneq I + \delta I\,,
\eal
with $I$ the total moment of inertia. 
Similarly, the kinetic energy reads
\beal
T &= \fract{M}{2}\,\sum_n\,\dot{\br}_n(\phi)\cdot\dot{\br}_n(\phi)\\
&= \fract{M}{2}\,\sum_n\,\big(\dot{\bx}_n + \dot{\phi}\,\bd{z}\wedge\br_n\big)\cdot
\big(\dot{\bx}_n + \dot{\phi}\,\bd{z}\wedge\br_n\big)\\
&=\fract{I_0}{2}\,\dot{\phi}^2 + J\,\dot{\phi} + \fract{M}{2}\,\dot{q}_{A_1}^2 + 
\fract{M}{2}\sum_{\ell\not=0}\,\dot{\bq}_\ell\cdot\dot{\bq}_{-\ell}\,.
\label{kinetic DN}
\eal 
By definition, the variables $p_\phi$, $p_{A_1}$ and $\bp_{-\ell}$ conjugate to 
$\phi$, $q_{A_1}$ and $\bq_\ell$, respectively, are obtained through 
\beal
&p_\phi = \fract{\partial T}{\partial\dot{\phi}} = \dot{\phi}\,I_0 + J \equiv \text{L}\,,\\
&p_{A_1} = \fract{\partial T}{\partial\dot{q}_{A_1}} = M\,\dot{q}_{A_1}\,,\\
&\bp_{-\ell} = \fract{\partial T}{\partial\dot{\bq}_{\ell}} =
M\,\dot{\bq}_{-\ell} + \dot{\phi}\;\fract{\partial J}{\partial\dot{\bq}_{\ell}}\;.
\eal
Conversely, 
\bealn
&\dot{\phi} = \fract{p_\phi-J}{I_0}\;,\qquad
\dot{q}_{A_1} = \fract{p_{A_1}}{M}\;,\\
&\dot{\bq}_\ell =\fract{1}{M}\,\bigg(\bp_{\ell}-\dot{\phi}\;\fract{\partial J}{\partial\dot{\bq}_{-\ell}}\bigg)\,.
\eal
It follows that 
\bealn
J &= \sum_{\ell\not=0}\,\dot{\bq}_\ell\cdot\fract{\partial J}{\partial\dot{\bq}_{\ell}}\\
&= \sum_{\ell\not=0}\,\bq_\ell\wedge\bp_{-\ell}\cdot\bd{z}
 - \fract{\dot{\phi}}{M}\,\sum_{\ell\not=0}\,\fract{\partial J}{\partial\dot{\bq}_{\ell}}
\cdot\fract{\partial J}{\partial\dot{\bq}_{-\ell}}\\
&\coloneq \text{L}_\text{vib} - \dot{\phi}\,M\,\sum_{\ell\not=0}\,
\bq_\ell\cdot\bq_{-\ell} = \text{L}_\text{vib} - \dot{\phi}\,\delta I\\
&= \text{L}_\text{vib} - \fract{\delta I}{I_0}\;\big(p_\phi-J\big)\,,
\eal
namely,
\bealn
J &= \fract{\;I_0\,\text{L}_\text{vib} -\delta I\,p_\phi\;}{I}\;,
\eal
and thus
\beal
\dot{\phi} &= \fract{\;p_\phi-\text{L}_\text{vib}\;}{I}\;.
\label{dot phi DN}
\eal
Therefore, if $V\big(\{\br_n\}\big)$ is the inter atomic potential, minimum at 
$\br_n=\bR_n$ and independent of $\phi$, the atomic Hamiltonian is defined through
\bealn
H_\text{at} &= p_\phi\,\dot{\phi} + p_{A_1}\,\dot{q}_{A_1} 
+ \sum_{\ell\not=0}\,\bp_{-\ell}\cdot\dot{\bq}_\ell - T + V\big(\{\br_n\}\big)\,,
\eal
which, using the above results, can be readily expressed in terms of conjugate variables as 
\beal
H_\text{at} &= \fract{\;\big(p_\phi- \text{L}_\text{vib}\big)^2\;}{2I}+ \fract{\;p_{A_1}^2}{2M} \\
&\qquad + \fract{1}{2M}\,\sum_{\ell\not=0}\,\bp_{\ell}\cdot\bp_{-\ell} + V\big(\{\br_n\}\big)\,.
\label{H DN}
\eal 
We emphasize that \eqn{H DN} is valid regardless of the form of the interatomic potential and can be straightforwardly quantized.


    
%


\begin{thebibliography}{24}%
  \makeatletter
  \providecommand \@ifxundefined [1]{%
   \@ifx{#1\undefined}
  }%
  \providecommand \@ifnum [1]{%
   \ifnum #1\expandafter \@firstoftwo
   \else \expandafter \@secondoftwo
   \fi
  }%
  \providecommand \@ifx [1]{%
   \ifx #1\expandafter \@firstoftwo
   \else \expandafter \@secondoftwo
   \fi
  }%
  \providecommand \natexlab [1]{#1}%
  \providecommand \enquote  [1]{``#1''}%
  \providecommand \bibnamefont  [1]{#1}%
  \providecommand \bibfnamefont [1]{#1}%
  \providecommand \citenamefont [1]{#1}%
  \providecommand \href@noop [0]{\@secondoftwo}%
  \providecommand \href [0]{\begingroup \@sanitize@url \@href}%
  \providecommand \@href[1]{\@@startlink{#1}\@@href}%
  \providecommand \@@href[1]{\endgroup#1\@@endlink}%
  \providecommand \@sanitize@url [0]{\catcode `\\12\catcode `\$12\catcode `\&12\catcode `\#12\catcode `\^12\catcode `\_12\catcode `\%12\relax}%
  \providecommand \@@startlink[1]{}%
  \providecommand \@@endlink[0]{}%
  \providecommand \url  [0]{\begingroup\@sanitize@url \@url }%
  \providecommand \@url [1]{\endgroup\@href {#1}{\urlprefix }}%
  \providecommand \urlprefix  [0]{URL }%
  \providecommand \Eprint [0]{\href }%
  \providecommand \doibase [0]{https://doi.org/}%
  \providecommand \selectlanguage [0]{\@gobble}%
  \providecommand \bibinfo  [0]{\@secondoftwo}%
  \providecommand \bibfield  [0]{\@secondoftwo}%
  \providecommand \translation [1]{[#1]}%
  \providecommand \BibitemOpen [0]{}%
  \providecommand \bibitemStop [0]{}%
  \providecommand \bibitemNoStop [0]{.\EOS\space}%
  \providecommand \EOS [0]{\spacefactor3000\relax}%
  \providecommand \BibitemShut  [1]{\csname bibitem#1\endcsname}%
  \let\auto@bib@innerbib\@empty
  \bibitem [{\citenamefont {Einstein}\ and\ \citenamefont {de~Haas}(1915)}]{Einstein-deHaas}%
    \BibitemOpen
    \bibfield  {author} {\bibinfo {author} {\bibfnamefont {A.}~\bibnamefont {Einstein}}\ and\ \bibinfo {author} {\bibfnamefont {W.}~\bibnamefont {de~Haas}},\ }\bibfield  {title} {\bibinfo {title} {{Experimenteller Nachweis der Ampereschen Molekularstr\"{o}me}},\ }\href@noop {} {\bibfield  {journal} {\bibinfo  {journal} {Verh. Dtsch. Phys. Ges.}\ }\textbf {\bibinfo {volume} {17}},\ \bibinfo {pages} {152} (\bibinfo {year} {1915})}\BibitemShut {NoStop}%
  \bibitem [{\citenamefont {Barnett}(1915)}]{Bernett-PR1915}%
    \BibitemOpen
    \bibfield  {author} {\bibinfo {author} {\bibfnamefont {S.~J.}\ \bibnamefont {Barnett}},\ }\bibfield  {title} {\bibinfo {title} {Magnetization by rotation},\ }\href {https://doi.org/10.1103/PhysRev.6.239} {\bibfield  {journal} {\bibinfo  {journal} {Phys. Rev.}\ }\textbf {\bibinfo {volume} {6}},\ \bibinfo {pages} {239} (\bibinfo {year} {1915})}\BibitemShut {NoStop}%
  \bibitem [{\citenamefont {Zhang}\ and\ \citenamefont {Niu}(2014)}]{Niu-PRL2014}%
    \BibitemOpen
    \bibfield  {author} {\bibinfo {author} {\bibfnamefont {L.}~\bibnamefont {Zhang}}\ and\ \bibinfo {author} {\bibfnamefont {Q.}~\bibnamefont {Niu}},\ }\bibfield  {title} {\bibinfo {title} {{Angular Momentum of Phonons and the Einstein--de Haas Effect}},\ }\href {https://doi.org/10.1103/PhysRevLett.112.085503} {\bibfield  {journal} {\bibinfo  {journal} {Phys. Rev. Lett.}\ }\textbf {\bibinfo {volume} {112}},\ \bibinfo {pages} {085503} (\bibinfo {year} {2014})}\BibitemShut {NoStop}%
  \bibitem [{\citenamefont {Garanin}\ and\ \citenamefont {Chudnovsky}(2015)}]{Garanin-PRB2015}%
    \BibitemOpen
    \bibfield  {author} {\bibinfo {author} {\bibfnamefont {D.~A.}\ \bibnamefont {Garanin}}\ and\ \bibinfo {author} {\bibfnamefont {E.~M.}\ \bibnamefont {Chudnovsky}},\ }\bibfield  {title} {\bibinfo {title} {Angular momentum in spin-phonon processes},\ }\href {https://doi.org/10.1103/PhysRevB.92.024421} {\bibfield  {journal} {\bibinfo  {journal} {Phys. Rev. B}\ }\textbf {\bibinfo {volume} {92}},\ \bibinfo {pages} {024421} (\bibinfo {year} {2015})}\BibitemShut {NoStop}%
  \bibitem [{\citenamefont {Dornes}\ \emph {et~al.}(2019)\citenamefont {Dornes}, \citenamefont {Acremann}, \citenamefont {Savoini}, \citenamefont {Kubli}, \citenamefont {Neugebauer}, \citenamefont {Abreu}, \citenamefont {Huber}, \citenamefont {Lantz}, \citenamefont {Vaz}, \citenamefont {Lemke}, \citenamefont {Bothschafter}, \citenamefont {Porer}, \citenamefont {Esposito}, \citenamefont {Rettig}, \citenamefont {Buzzi}, \citenamefont {Alberca}, \citenamefont {Windsor}, \citenamefont {Beaud}, \citenamefont {Staub}, \citenamefont {Zhu}, \citenamefont {Song}, \citenamefont {Glownia},\ and\ \citenamefont {Johnson}}]{Dornes-Nature2019}%
    \BibitemOpen
    \bibfield  {author} {\bibinfo {author} {\bibfnamefont {C.}~\bibnamefont {Dornes}}, \bibinfo {author} {\bibfnamefont {Y.}~\bibnamefont {Acremann}}, \bibinfo {author} {\bibfnamefont {M.}~\bibnamefont {Savoini}}, \bibinfo {author} {\bibfnamefont {M.}~\bibnamefont {Kubli}}, \bibinfo {author} {\bibfnamefont {M.~J.}\ \bibnamefont {Neugebauer}}, \bibinfo {author} {\bibfnamefont {E.}~\bibnamefont {Abreu}}, \bibinfo {author} {\bibfnamefont {L.}~\bibnamefont {Huber}}, \bibinfo {author} {\bibfnamefont {G.}~\bibnamefont {Lantz}}, \bibinfo {author} {\bibfnamefont {C.~A.~F.}\ \bibnamefont {Vaz}}, \bibinfo {author} {\bibfnamefont {H.}~\bibnamefont {Lemke}}, \bibinfo {author} {\bibfnamefont {E.~M.}\ \bibnamefont {Bothschafter}}, \bibinfo {author} {\bibfnamefont {M.}~\bibnamefont {Porer}}, \bibinfo {author} {\bibfnamefont {V.}~\bibnamefont {Esposito}}, \bibinfo {author} {\bibfnamefont {L.}~\bibnamefont {Rettig}}, \bibinfo {author} {\bibfnamefont {M.}~\bibnamefont {Buzzi}}, \bibinfo {author} {\bibfnamefont {A.}~\bibnamefont {Alberca}}, \bibinfo {author} {\bibfnamefont {Y.~W.}\ \bibnamefont {Windsor}}, \bibinfo {author} {\bibfnamefont {P.}~\bibnamefont {Beaud}}, \bibinfo {author} {\bibfnamefont {U.}~\bibnamefont {Staub}}, \bibinfo {author} {\bibfnamefont {D.}~\bibnamefont {Zhu}}, \bibinfo {author} {\bibfnamefont {S.}~\bibnamefont {Song}}, \bibinfo {author} {\bibfnamefont {J.~M.}\ \bibnamefont {Glownia}},\ and\ \bibinfo {author} {\bibfnamefont {S.~L.}\ \bibnamefont {Johnson}},\ }\bibfield  {title} {\bibinfo {title} {{The ultrafast Einstein--de Haas effect}},\ }\href {https://doi.org/10.1038/s41586-018-0822-7} {\bibfield  {journal} {\bibinfo  {journal} {Nature}\ }\textbf {\bibinfo {volume} {565}},\ \bibinfo {pages} {209} (\bibinfo {year} {2019})}\BibitemShut {NoStop}%
  \bibitem [{\citenamefont {R\"uckriegel}\ \emph {et~al.}(2020)\citenamefont {R\"uckriegel}, \citenamefont {Streib}, \citenamefont {Bauer},\ and\ \citenamefont {Duine}}]{Rembert-PRB2020}%
    \BibitemOpen
    \bibfield  {author} {\bibinfo {author} {\bibfnamefont {A.}~\bibnamefont {R\"uckriegel}}, \bibinfo {author} {\bibfnamefont {S.}~\bibnamefont {Streib}}, \bibinfo {author} {\bibfnamefont {G.~E.~W.}\ \bibnamefont {Bauer}},\ and\ \bibinfo {author} {\bibfnamefont {R.~A.}\ \bibnamefont {Duine}},\ }\bibfield  {title} {\bibinfo {title} {Angular momentum conservation and phonon spin in magnetic insulators},\ }\href {https://doi.org/10.1103/PhysRevB.101.104402} {\bibfield  {journal} {\bibinfo  {journal} {Phys. Rev. B}\ }\textbf {\bibinfo {volume} {101}},\ \bibinfo {pages} {104402} (\bibinfo {year} {2020})}\BibitemShut {NoStop}%
  \bibitem [{\citenamefont {Garanin}\ and\ \citenamefont {Chudnovsky}(2021)}]{Garanin-PRB2021}%
    \BibitemOpen
    \bibfield  {author} {\bibinfo {author} {\bibfnamefont {D.~A.}\ \bibnamefont {Garanin}}\ and\ \bibinfo {author} {\bibfnamefont {E.~M.}\ \bibnamefont {Chudnovsky}},\ }\bibfield  {title} {\bibinfo {title} {Conservation of angular momentum in an elastic medium with spins},\ }\href {https://doi.org/10.1103/PhysRevB.103.L100412} {\bibfield  {journal} {\bibinfo  {journal} {Phys. Rev. B}\ }\textbf {\bibinfo {volume} {103}},\ \bibinfo {pages} {L100412} (\bibinfo {year} {2021})}\BibitemShut {NoStop}%
  \bibitem [{\citenamefont {Tauchert}\ \emph {et~al.}(2022)\citenamefont {Tauchert}, \citenamefont {Volkov}, \citenamefont {Ehberger}, \citenamefont {Kazenwadel}, \citenamefont {Evers}, \citenamefont {Lange}, \citenamefont {Donges}, \citenamefont {Book}, \citenamefont {Kreuzpaintner}, \citenamefont {Nowak},\ and\ \citenamefont {Baum}}]{Tauchert-Nature2022}%
    \BibitemOpen
    \bibfield  {author} {\bibinfo {author} {\bibfnamefont {S.~R.}\ \bibnamefont {Tauchert}}, \bibinfo {author} {\bibfnamefont {M.}~\bibnamefont {Volkov}}, \bibinfo {author} {\bibfnamefont {D.}~\bibnamefont {Ehberger}}, \bibinfo {author} {\bibfnamefont {D.}~\bibnamefont {Kazenwadel}}, \bibinfo {author} {\bibfnamefont {M.}~\bibnamefont {Evers}}, \bibinfo {author} {\bibfnamefont {H.}~\bibnamefont {Lange}}, \bibinfo {author} {\bibfnamefont {A.}~\bibnamefont {Donges}}, \bibinfo {author} {\bibfnamefont {A.}~\bibnamefont {Book}}, \bibinfo {author} {\bibfnamefont {W.}~\bibnamefont {Kreuzpaintner}}, \bibinfo {author} {\bibfnamefont {U.}~\bibnamefont {Nowak}},\ and\ \bibinfo {author} {\bibfnamefont {P.}~\bibnamefont {Baum}},\ }\bibfield  {title} {\bibinfo {title} {Polarized phonons carry angular momentum in ultrafast demagnetization},\ }\href {https://doi.org/10.1038/s41586-021-04306-4} {\bibfield  {journal} {\bibinfo  {journal} {Nature}\ }\textbf {\bibinfo {volume} {602}},\ \bibinfo {pages} {73} (\bibinfo {year} {2022})}\BibitemShut {NoStop}%
  \bibitem [{\citenamefont {Zhang}\ and\ \citenamefont {Niu}(2015)}]{Niu-PRL2015}%
    \BibitemOpen
    \bibfield  {author} {\bibinfo {author} {\bibfnamefont {L.}~\bibnamefont {Zhang}}\ and\ \bibinfo {author} {\bibfnamefont {Q.}~\bibnamefont {Niu}},\ }\bibfield  {title} {\bibinfo {title} {Chiral phonons at high-symmetry points in monolayer hexagonal lattices},\ }\href {https://doi.org/10.1103/PhysRevLett.115.115502} {\bibfield  {journal} {\bibinfo  {journal} {Phys. Rev. Lett.}\ }\textbf {\bibinfo {volume} {115}},\ \bibinfo {pages} {115502} (\bibinfo {year} {2015})}\BibitemShut {NoStop}%
  \bibitem [{\citenamefont {Hamada}\ \emph {et~al.}(2018)\citenamefont {Hamada}, \citenamefont {Minamitani}, \citenamefont {Hirayama},\ and\ \citenamefont {Murakami}}]{Hamada-PRL2018}%
    \BibitemOpen
    \bibfield  {author} {\bibinfo {author} {\bibfnamefont {M.}~\bibnamefont {Hamada}}, \bibinfo {author} {\bibfnamefont {E.}~\bibnamefont {Minamitani}}, \bibinfo {author} {\bibfnamefont {M.}~\bibnamefont {Hirayama}},\ and\ \bibinfo {author} {\bibfnamefont {S.}~\bibnamefont {Murakami}},\ }\bibfield  {title} {\bibinfo {title} {Phonon angular momentum induced by the temperature gradient},\ }\href {https://doi.org/10.1103/PhysRevLett.121.175301} {\bibfield  {journal} {\bibinfo  {journal} {Phys. Rev. Lett.}\ }\textbf {\bibinfo {volume} {121}},\ \bibinfo {pages} {175301} (\bibinfo {year} {2018})}\BibitemShut {NoStop}%
  \bibitem [{\citenamefont {Zhong}\ \emph {et~al.}(2023)\citenamefont {Zhong}, \citenamefont {Sun}, \citenamefont {Pan}, \citenamefont {Wang}, \citenamefont {Xu}, \citenamefont {Zhang},\ and\ \citenamefont {Zhou}}]{Zhong-PRB2023}%
    \BibitemOpen
    \bibfield  {author} {\bibinfo {author} {\bibfnamefont {J.}~\bibnamefont {Zhong}}, \bibinfo {author} {\bibfnamefont {H.}~\bibnamefont {Sun}}, \bibinfo {author} {\bibfnamefont {Y.}~\bibnamefont {Pan}}, \bibinfo {author} {\bibfnamefont {Z.}~\bibnamefont {Wang}}, \bibinfo {author} {\bibfnamefont {X.}~\bibnamefont {Xu}}, \bibinfo {author} {\bibfnamefont {L.}~\bibnamefont {Zhang}},\ and\ \bibinfo {author} {\bibfnamefont {J.}~\bibnamefont {Zhou}},\ }\bibfield  {title} {\bibinfo {title} {{Abnormal phonon angular momentum due to off-diagonal elements in the density matrix induced by a temperature gradient}},\ }\href {https://doi.org/10.1103/PhysRevB.107.125147} {\bibfield  {journal} {\bibinfo  {journal} {Phys. Rev. B}\ }\textbf {\bibinfo {volume} {107}},\ \bibinfo {pages} {125147} (\bibinfo {year} {2023})}\BibitemShut {NoStop}%
  \bibitem [{\citenamefont {Zhang}\ \emph {et~al.}(2024)\citenamefont {Zhang}, \citenamefont {Peshcherenko}, \citenamefont {Yang}, \citenamefont {Ward}, \citenamefont {Raghuvanshi}, \citenamefont {Lindsay}, \citenamefont {Felser}, \citenamefont {Zhang}, \citenamefont {Yan},\ and\ \citenamefont {Miao}}]{zhang-ArXiv2024}%
    \BibitemOpen
    \bibfield  {author} {\bibinfo {author} {\bibfnamefont {H.}~\bibnamefont {Zhang}}, \bibinfo {author} {\bibfnamefont {N.}~\bibnamefont {Peshcherenko}}, \bibinfo {author} {\bibfnamefont {F.}~\bibnamefont {Yang}}, \bibinfo {author} {\bibfnamefont {T.~Z.}\ \bibnamefont {Ward}}, \bibinfo {author} {\bibfnamefont {P.}~\bibnamefont {Raghuvanshi}}, \bibinfo {author} {\bibfnamefont {L.}~\bibnamefont {Lindsay}}, \bibinfo {author} {\bibfnamefont {C.}~\bibnamefont {Felser}}, \bibinfo {author} {\bibfnamefont {Y.}~\bibnamefont {Zhang}}, \bibinfo {author} {\bibfnamefont {J.~Q.}\ \bibnamefont {Yan}},\ and\ \bibinfo {author} {\bibfnamefont {H.}~\bibnamefont {Miao}},\ }\href {https://arxiv.org/abs/2409.13462} {\bibinfo {title} {Observation of phonon angular momentum}} (\bibinfo {year} {2024}),\ \Eprint {https://arxiv.org/abs/2409.13462} {arXiv:2409.13462 [cond-mat.str-el]} \BibitemShut {NoStop}%
  \bibitem [{\citenamefont {Wang}\ \emph {et~al.}(2024)\citenamefont {Wang}, \citenamefont {Sun}, \citenamefont {Li},\ and\ \citenamefont {Zhang}}]{Zhang-NanoLett2024}%
    \BibitemOpen
    \bibfield  {author} {\bibinfo {author} {\bibfnamefont {T.}~\bibnamefont {Wang}}, \bibinfo {author} {\bibfnamefont {H.}~\bibnamefont {Sun}}, \bibinfo {author} {\bibfnamefont {X.}~\bibnamefont {Li}},\ and\ \bibinfo {author} {\bibfnamefont {L.}~\bibnamefont {Zhang}},\ }\bibfield  {title} {\bibinfo {title} {Chiral phonons: Prediction, verification, and application},\ }\href {https://doi.org/10.1021/acs.nanolett.4c00606} {\bibfield  {journal} {\bibinfo  {journal} {Nano Letters}\ }\textbf {\bibinfo {volume} {24}},\ \bibinfo {pages} {4311} (\bibinfo {year} {2024})}\BibitemShut {NoStop}%
  \bibitem [{\citenamefont {Streib}(2021)}]{Streib-PRB2021}%
    \BibitemOpen
    \bibfield  {author} {\bibinfo {author} {\bibfnamefont {S.}~\bibnamefont {Streib}},\ }\bibfield  {title} {\bibinfo {title} {Difference between angular momentum and pseudoangular momentum},\ }\href {https://doi.org/10.1103/PhysRevB.103.L100409} {\bibfield  {journal} {\bibinfo  {journal} {Phys. Rev. B}\ }\textbf {\bibinfo {volume} {103}},\ \bibinfo {pages} {L100409} (\bibinfo {year} {2021})}\BibitemShut {NoStop}%
  \bibitem [{\citenamefont {Chaudhary}\ \emph {et~al.}(2024)\citenamefont {Chaudhary}, \citenamefont {Juraschek}, \citenamefont {Rodriguez-Vega},\ and\ \citenamefont {Fiete}}]{Gregory-PRB2024}%
    \BibitemOpen
    \bibfield  {author} {\bibinfo {author} {\bibfnamefont {S.}~\bibnamefont {Chaudhary}}, \bibinfo {author} {\bibfnamefont {D.~M.}\ \bibnamefont {Juraschek}}, \bibinfo {author} {\bibfnamefont {M.}~\bibnamefont {Rodriguez-Vega}},\ and\ \bibinfo {author} {\bibfnamefont {G.~A.}\ \bibnamefont {Fiete}},\ }\bibfield  {title} {\bibinfo {title} {Giant effective magnetic moments of chiral phonons from orbit-lattice coupling},\ }\href {https://doi.org/10.1103/PhysRevB.110.094401} {\bibfield  {journal} {\bibinfo  {journal} {Phys. Rev. B}\ }\textbf {\bibinfo {volume} {110}},\ \bibinfo {pages} {094401} (\bibinfo {year} {2024})}\BibitemShut {NoStop}%
  \bibitem [{\citenamefont {Rice}\ and\ \citenamefont {Choi}(1992)}]{Rice&Choi}%
    \BibitemOpen
    \bibfield  {author} {\bibinfo {author} {\bibfnamefont {M.~J.}\ \bibnamefont {Rice}}\ and\ \bibinfo {author} {\bibfnamefont {H.-Y.}\ \bibnamefont {Choi}},\ }\bibfield  {title} {\bibinfo {title} {Charged-phonon absorption in doped ${\mathrm{c}}_{60}$},\ }\href {https://doi.org/10.1103/PhysRevB.45.10173} {\bibfield  {journal} {\bibinfo  {journal} {Phys. Rev. B}\ }\textbf {\bibinfo {volume} {45}},\ \bibinfo {pages} {10173} (\bibinfo {year} {1992})}\BibitemShut {NoStop}%
  \bibitem [{\citenamefont {Pak}\ \emph {et~al.}(1997)\citenamefont {Pak}, \citenamefont {Sibert},\ and\ \citenamefont {Woods}}]{Pak-JCP1997}%
    \BibitemOpen
    \bibfield  {author} {\bibinfo {author} {\bibfnamefont {Y.}~\bibnamefont {Pak}}, \bibinfo {author} {\bibfnamefont {I.}~\bibnamefont {Sibert}, \bibfnamefont {Edwin~L.}},\ and\ \bibinfo {author} {\bibfnamefont {R.~C.}\ \bibnamefont {Woods}},\ }\bibfield  {title} {\bibinfo {title} {Coupled cluster anharmonic force fields, spectroscopic constants, and vibrational energies of {AIF}$_3$ and {SiF}$_3^+$},\ }\href {https://doi.org/10.1063/1.474613} {\bibfield  {journal} {\bibinfo  {journal} {The Journal of Chemical Physics}\ }\textbf {\bibinfo {volume} {107}},\ \bibinfo {pages} {1717} (\bibinfo {year} {1997})},\ \Eprint {https://arxiv.org/abs/https://pubs.aip.org/aip/jcp/article-pdf/107/6/1717/19218415/1717\_1\_online.pdf} {https://pubs.aip.org/aip/jcp/article-pdf/107/6/1717/19218415/1717\_1\_online.pdf} \BibitemShut {NoStop}%
  \bibitem [{\citenamefont {Nejad}\ and\ \citenamefont {Crittenden}(2020)}]{Nejad-PCCP2020}%
    \BibitemOpen
    \bibfield  {author} {\bibinfo {author} {\bibfnamefont {A.}~\bibnamefont {Nejad}}\ and\ \bibinfo {author} {\bibfnamefont {D.~L.}\ \bibnamefont {Crittenden}},\ }\bibfield  {title} {\bibinfo {title} {On the separability of large-amplitude motions in anharmonic frequency calculations},\ }\href {https://doi.org/10.1039/D0CP03515G} {\bibfield  {journal} {\bibinfo  {journal} {Phys. Chem. Chem. Phys.}\ }\textbf {\bibinfo {volume} {22}},\ \bibinfo {pages} {20588} (\bibinfo {year} {2020})}\BibitemShut {NoStop}%
  \bibitem [{\citenamefont {Solomonik}\ and\ \citenamefont {Marochko}(2000)}]{Solomik-JSC2000}%
    \BibitemOpen
    \bibfield  {author} {\bibinfo {author} {\bibfnamefont {V.~G.}\ \bibnamefont {Solomonik}}\ and\ \bibinfo {author} {\bibfnamefont {O.~Y.}\ \bibnamefont {Marochko}},\ }\bibfield  {title} {\bibinfo {title} {{Molecular Structures and Vibrational Spectra of ScF3, YF3, and LaF3 as Calculated by the CISD+Q Method}},\ }\href {https://doi.org/10.1023/A:1004834730711} {\bibfield  {journal} {\bibinfo  {journal} {Journal of Structural Chemistry}\ }\textbf {\bibinfo {volume} {41}},\ \bibinfo {pages} {725} (\bibinfo {year} {2000})}\BibitemShut {NoStop}%
  \bibitem [{\citenamefont {Zwanziger}\ and\ \citenamefont {Grant}(1987)}]{Zwanziger&Grant-1987}%
    \BibitemOpen
    \bibfield  {author} {\bibinfo {author} {\bibfnamefont {J.~W.}\ \bibnamefont {Zwanziger}}\ and\ \bibinfo {author} {\bibfnamefont {E.~R.}\ \bibnamefont {Grant}},\ }\bibfield  {title} {\bibinfo {title} {{Topological phase in molecular bound states: Application to the {$E\times e$} system}},\ }\href {https://doi.org/10.1063/1.453083} {\bibfield  {journal} {\bibinfo  {journal} {The Journal of Chemical Physics}\ }\textbf {\bibinfo {volume} {87}},\ \bibinfo {pages} {2954} (\bibinfo {year} {1987})},\ \Eprint {https://arxiv.org/abs/https://pubs.aip.org/aip/jcp/article-pdf/87/5/2954/18965948/2954\_1\_online.pdf} {https://pubs.aip.org/aip/jcp/article-pdf/87/5/2954/18965948/2954\_1\_online.pdf} \BibitemShut {NoStop}%
  \bibitem [{\citenamefont {Maslen}\ \emph {et~al.}(1992)\citenamefont {Maslen}, \citenamefont {Handy}, \citenamefont {Amos},\ and\ \citenamefont {Jayatilaka}}]{Maslen-JCP1992}%
    \BibitemOpen
    \bibfield  {author} {\bibinfo {author} {\bibfnamefont {P.~E.}\ \bibnamefont {Maslen}}, \bibinfo {author} {\bibfnamefont {N.~C.}\ \bibnamefont {Handy}}, \bibinfo {author} {\bibfnamefont {R.~D.}\ \bibnamefont {Amos}},\ and\ \bibinfo {author} {\bibfnamefont {D.}~\bibnamefont {Jayatilaka}},\ }\bibfield  {title} {\bibinfo {title} {{Higher analytic derivatives. IV. Anharmonic effects in the benzene spectrum}},\ }\href {https://doi.org/10.1063/1.463926} {\bibfield  {journal} {\bibinfo  {journal} {The Journal of Chemical Physics}\ }\textbf {\bibinfo {volume} {97}},\ \bibinfo {pages} {4233} (\bibinfo {year} {1992})},\ \Eprint {https://arxiv.org/abs/https://pubs.aip.org/aip/jcp/article-pdf/97/6/4233/19001899/4233\_1\_online.pdf} {https://pubs.aip.org/aip/jcp/article-pdf/97/6/4233/19001899/4233\_1\_online.pdf} \BibitemShut {NoStop}%
  \bibitem [{\citenamefont {Medina}\ \emph {et~al.}(2015)\citenamefont {Medina}, \citenamefont {Avilés},\ and\ \citenamefont {Tapia}}]{Medina03062015}%
    \BibitemOpen
    \bibfield  {author} {\bibinfo {author} {\bibfnamefont {J.}~\bibnamefont {Medina}}, \bibinfo {author} {\bibfnamefont {F.}~\bibnamefont {Avilés}},\ and\ \bibinfo {author} {\bibfnamefont {A.}~\bibnamefont {Tapia}},\ }\bibfield  {title} {\bibinfo {title} {The bond force constants of graphene and benzene calculated by density functional theory},\ }\href {https://doi.org/10.1080/00268976.2014.986241} {\bibfield  {journal} {\bibinfo  {journal} {Molecular Physics}\ }\textbf {\bibinfo {volume} {113}},\ \bibinfo {pages} {1297} (\bibinfo {year} {2015})},\ \Eprint {https://arxiv.org/abs/https://doi.org/10.1080/00268976.2014.986241} {https://doi.org/10.1080/00268976.2014.986241} \BibitemShut {NoStop}%
  \bibitem [{\citenamefont {Preuss}\ and\ \citenamefont {Bechstedt}(2006)}]{fononi_benzene}%
    \BibitemOpen
    \bibfield  {author} {\bibinfo {author} {\bibfnamefont {M.}~\bibnamefont {Preuss}}\ and\ \bibinfo {author} {\bibfnamefont {F.}~\bibnamefont {Bechstedt}},\ }\bibfield  {title} {\bibinfo {title} {Vibrational spectra of ammonia, benzene, and benzene adsorbed on {Si} (110) by \textit{first principles} calculations with periodic boundary conditions},\ }\href {https://doi.org/10.1103/PhysRevB.73.155413} {\bibfield  {journal} {\bibinfo  {journal} {Phys. Rev. B}\ }\textbf {\bibinfo {volume} {73}},\ \bibinfo {pages} {155413} (\bibinfo {year} {2006})}\BibitemShut {NoStop}%
  \bibitem [{\citenamefont {Bertie}\ and\ \citenamefont {Keefe}(2004)}]{dipolo_benzene}%
    \BibitemOpen
    \bibfield  {author} {\bibinfo {author} {\bibfnamefont {J.~E.}\ \bibnamefont {Bertie}}\ and\ \bibinfo {author} {\bibfnamefont {C.~D.}\ \bibnamefont {Keefe}},\ }\bibfield  {title} {\bibinfo {title} {{Infrared intensities of liquids XXIV: optical constants of liquid benzene‑h$_6$ at 25\,$^\circ$C extended to 11.5\,cm$^{-1}$ and molar polarizabilities and integrated intensities of benzene‑h$_6$ between 6200 and 11.5\,cm$^{-1}$}},\ }\href {https://doi.org/10.1016/j.molstruc.2003.11.002} {\bibfield  {journal} {\bibinfo  {journal} {Journal of Molecular Structure}\ }\textbf {\bibinfo {volume} {695--696}},\ \bibinfo {pages} {39} (\bibinfo {year} {2004})},\ \bibinfo {note} {{Winnewisser Special Issue}}\BibitemShut {NoStop}%
  \end{thebibliography}
\end{document}